\documentclass{emulateapj}
\usepackage{natbib}
\usepackage{apjfonts}

\newcommand{\NEP}[1]{\mbox{NEP}_{\mbox{\scriptsize #1}}}

\newcommand{\NET}[1]{\mbox{NET}_{\mbox{\scriptsize #1}}}

\newcommand{\trans}{{\mathcal T}}
\newcommand{\Tcmb}{T_{\mbox{\tiny CMB}}}
\newcommand{\istk}{{\mathcal{I}}}
\newcommand{\qstk}{{\mathcal{Q}}}
\newcommand{\ustk}{{\mathcal{U}}}

\newcommand{\ns}[1]{ {\mbox{\scriptsize #1}} }

\newcommand{\tn}[1]{\tablenotemark{#1}}
\newcommand{\clempaper}{the Data Paper}
\newcommand{\be}{\begin{equation}}
\newcommand{\ee}{\end{equation}}

\slugcomment{To appear in the Astrophysical Journal}

\shorttitle{The QUaD instrument}
\shortauthors{Hinderks et al.}

\begin{document}

\title{QUaD: A High-Resolution Cosmic Microwave Background Polarimeter}

\author{J.\,R.\,Hinderks\altaffilmark{1,2}, P.\,Ade\altaffilmark{3}, J.\,Bock\altaffilmark{4,5},
        M.\,Bowden\altaffilmark{1}, M.\,L.\,Brown\altaffilmark{6,7},
        G.\,Cahill\altaffilmark{8}, J.\,E.\,Carlstrom\altaffilmark{9},
        P.\,G.\,Castro\altaffilmark{6,10}, S.\,Church\altaffilmark{1},
        T.\,Culverhouse\altaffilmark{9}, R.\,Friedman\altaffilmark{9},
        K.\,Ganga\altaffilmark{11}, W.\,K.\,Gear\altaffilmark{3},
        S.\,Gupta\altaffilmark{3},
        J.\,Harris\altaffilmark{3},  V.\,Haynes\altaffilmark{3,12},
        B.\,G.\,Keating\altaffilmark{13}, J.\,Kovac\altaffilmark{4,5},
        E.\,Kirby\altaffilmark{1},
        A.\,E.\,Lange\altaffilmark{5}, E.\,Leitch\altaffilmark{4,5},
        O.\,E.\,Mallie\altaffilmark{3},
        S.\,Melhuish\altaffilmark{3,12}, Y.\,Memari\altaffilmark{6},
        J.\,A.\,Murphy\altaffilmark{8}, A.\,Orlando\altaffilmark{3,5},
        R.\,Schwarz\altaffilmark{9}, C.\,O'\,Sullivan\altaffilmark{8},
        L.\,Piccirillo\altaffilmark{3,12}, C.\,Pryke\altaffilmark{9},
        N.\,Rajguru\altaffilmark{14}, B.\,Rusholme\altaffilmark{1,4},
        A.\,N.\,Taylor\altaffilmark{6}, K.\,L.\,Thompson\altaffilmark{1},
        C.\,Tucker\altaffilmark{3}, A.\,H.\,Turner\altaffilmark{3},
        E.\,Y.\,S.\,Wu\altaffilmark{1} and M.\,Zemcov\altaffilmark{4,5}}
\altaffiltext{1}{Kavli Institute for Particle Astrophysics and Cosmology and Department of Physics, Stanford University,
382 Via Pueblo Mall, Stanford, CA 94305, USA.}
\altaffiltext{2}{\emph{Current address:} NASA Goddard Space Flight Center, 8800 Greenbelt
  Road, Greenbelt, Maryland 20771, USA.}
\altaffiltext{3}{School of Physics and Astronomy, Cardiff University,
  Queen's Buildings, The Parade, Cardiff CF24 3AA, UK.}
\altaffiltext{4}{Jet Propulsion Laboratory, 4800 Oak Grove Dr.,
  Pasadena, CA 91109, USA.}
\altaffiltext{5}{California Institute of Technology, Pasadena, CA
  91125, USA.}
\altaffiltext{6}{Institute for Astronomy, University of Edinburgh,
  Royal Observatory, Blackford Hill, Edinburgh EH9 3HJ, UK.}
\altaffiltext{7}{\emph{Current address:} Cavendish Laboratory, University of Cambridge,
  J.J. Thomson Avenue, Cambridge CB3 OHE, UK.}
\altaffiltext{8}{Experimental Physics, National University of Ireland,
  Maynooth, Ireland.}
\altaffiltext{9}{Kavli Institute for Cosmological Physics,
  Department of Astronomy \& Astrophysics, University of Chicago,
  5640 South Ellis Avenue, Chicago, IL 60637, USA.}
\altaffiltext{10}{\emph{Current address:} CENTRA, Departamento de F\'isica, Edif\'icio Ci\^encia,
  Instituto Superior T\'ecnico, Universidade Tecnica de Lisboa,
  Av. Rovisco Pais 1, 1049-001 Lisboa, Portugal.}
\altaffiltext{11}{Laboratoire APC/CNRS; B\^atiment Condorcet; 10, rue Alice Domon et L\'eonie Duquet; 75205 Paris Cedex 13; France.}
\altaffiltext{12}{\emph{Current address:} School of Physics and Astronomy, University of
  Manchester, Manchester M13 9PL, UK.}
\altaffiltext{13}{Center for Astrophysics and Space Sciences, University of California,
San Diego, 9500 Gilman Drive, La Jolla, CA 92093, USA.}
\altaffiltext{14}{\emph{Current address:} Department of Physics and Astronomy, University
  College London, Gower Street, London WC1E 6BT, UK.}

\begin{abstract}
We describe the QUaD experiment, a millimeter-wavelength polarimeter designed to observe the Cosmic Microwave Background (CMB) from a site at the South Pole.
The experiment comprises a 2.64~m Cassegrain telescope equipped with a cryogenically cooled receiver containing an array of 62 polarization-sensitive bolometers.
The focal plane contains pixels at two different frequency bands, 100\,GHz and 150\,GHz, with angular resolutions of $5^\prime$ and $3.5^\prime$, respectively.
The high angular resolution allows observation of CMB temperature and polarization anisotropies over a wide range of scales.
The instrument commenced operation in early 2005 and collected science data during three successive Austral winter seasons of observation.
\end{abstract}

\keywords{cosmic microwave background --- polarization --- instrumentation \vspace{-0.075in}}

\section{Introduction} \label{sec:introduction}
The Cosmic Microwave Background (CMB) remains a key tool for
understanding the origin and evolution of the universe. Thompson
scattering from quadrupole anisotropies at the surface of last
scattering polarizes the CMB at the level of 10\%. The resulting
polarization pattern on the sky can be mathematically decomposed into
even-parity E modes and odd-parity B modes \citep{zaldarriaga1997}. The
E mode signal, which has been detected by a number of experiments
\citep{readhead2004, leitch2005, montroy06, page07, wu07, 2008CAPMAP},
is dominated by scalar perturbations (density fluctuations) in the
early universe. A B-mode signal has yet to be detected but could be
generated by gravitational waves in the early universe or lensing
\mbox{of E-modes by intervening structure.}

This paper describes QUaD,\footnote{QUaD stands for QUEST (QU
Extragalactic Survey Telescope) at DASI (Degree Angular Scale
Interferometer).} a polarimeter designed to observe the CMB.
QUaD comprises a bolometric receiver located on a 2.64\,m telescope near the geographic South
Pole.
The QUaD receiver contains a focal plane array of 31 pixels, each composed of a corrugated feed horn
and a pair of orthogonal polarization-sensitive bolometers (PSBs).
Each pixel simultaneously measures both temperature and one linear polarization Stokes parameter.
The pixels are divided between two observing frequencies
with 12 at 100\,GHz and 19 at 150\,GHz.
The angular resolution is $5.0^\prime$ at 100\,GHz and $3.5^\prime$ at 150\,GHz, and the instantaneous field of view is $1.5^\circ$.
First light was in February 2005, and three seasons of Austral winter observations were completed before
the instrument was decommissioned in late 2007.
Results from the first season's data are presented in \citet{quad_season1}.
Results from the second and third seasons' data
are presented in \citet{Pryke2008}, a companion paper
that will be referred to in this work as \clempaper.

This paper is organized as follows: Sections~\ref{sec:mount}
and~\ref{sec:optics} review the observing site, telescope mount, and optics.
Sections~\ref{sec:receiver}, \ref{sec:cryo} and \ref{sec:electronics}
describe the focal plane, cryogenics, and readout electronics.
Section~\ref{sec:performance} presents the performance of the
instrument as measured in the laboratory and the field.
Section~\ref{sec:interference} describes the measures taken
to mitigate interfering signals.
Section~\ref{sec:calibration} describes the calibration procedures for
the instrument.
Section~\ref{sec:sensitivity} discusses the instrument sensitivity.

\section{Observing Site and Telescope Mount} \label{sec:mount}
\begin{figure}[t]
    \centering
\includegraphics[width=3.25in]{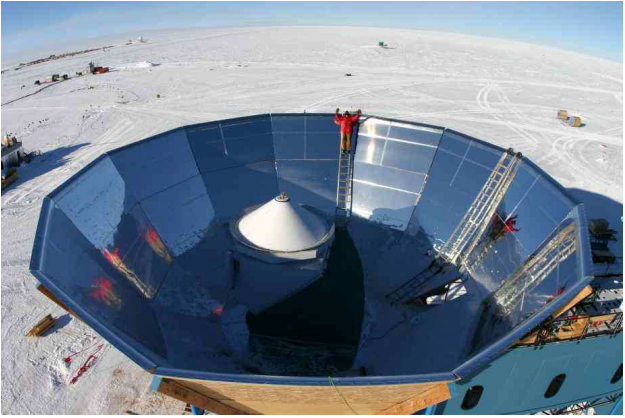}
    \caption{The QUaD telescope within the reflective ground shield.}\label{fig:quad_photograph}
\end{figure}

\begin{figure}
    \centering
\includegraphics[width=3.25in]{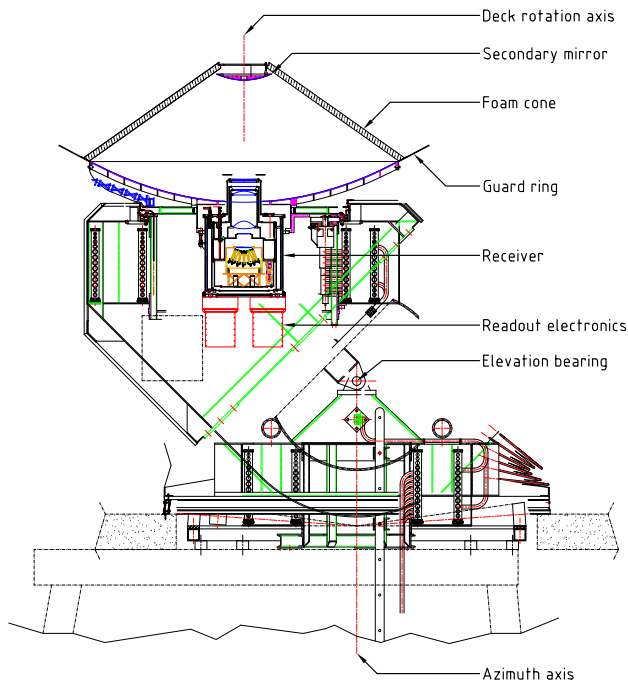}
    \caption{Schematic of the telescope and receiver.}\label{fig:quad_general_assembly}
\end{figure}

QUaD is sited at the Martin A. Pomerantz Observatory (MAPO), part of
the Amundsen-Scott Station, 0.7\,km from the geographic South Pole. The
Observatory is atop the Antarctic Plateau at a physical elevation of
2800\,m and at an equivalent pressure elevation in excess of 3200\,m.
This is recognized as a premier site for ground-based millimeter wave observations
\citep{lane98,lay2000,peterson2003,bussmann}.
The low temperature freezes out much of the remaining precipitable
water vapor above the telescope, reducing emission and absorption at
millimeter wavelengths. The temperature and atmosphere are stable over
long periods of time with minimal diurnal variation during the six-month
darkness of the Austral winter. The site also affords unchanging
access to the Southern Celestial Hemisphere, allowing deep and
continuous integration on the target region with rigorous control of
ground-spill systematics.

Figure~\ref{fig:quad_photograph} shows the observatory and Figure~\ref{fig:quad_general_assembly} shows a schematic of the experiment.
The telescope is located on the mount previously used by the DASI
experiment \citep{leitch_nature,leitch_2002}.
The mount is an altitude-azimuth design, with an additional axis of rotation that
allows the entire telescope and receiver to be rotated about the
optical axis of the instrument.
Known as ``deck'' rotation, this allows each
detector to view the sky at multiple polarization angles, despite the
fixed parallactic angle of sources viewed from the Polar location.
Observations are made by scanning the telescope in azimuth,
stepping in elevation between scans.

The telescope mount is isolated from wind loading and other sources of vibration
by being situated on the inner of two concentric, mechanically isolated
towers.
The outer tower supports the reflective ground shield and is connected to the
observatory building.
The receiver, readout electronics, and cryogen lines for filling are accessed from a heated room
below the mount.
This arrangement minimizes outdoor activity,
which is difficult in winter due to darkness and
extreme cold (as low as $-80^\circ$\,C ambient).

\section{Optics} \label{sec:optics}
\begin{figure}
    \begin{center}
\includegraphics[width=2.4in,clip]{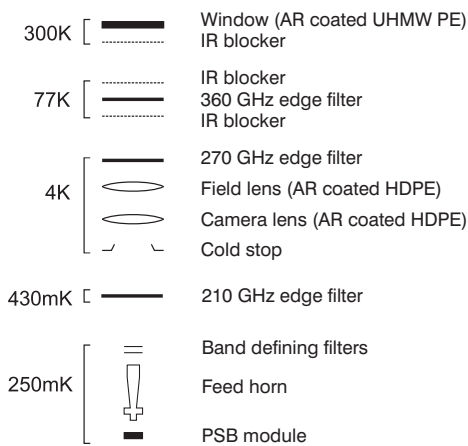}
    \end{center}
    \caption{The receiver optics and filter chain.  The window is made from anti-reflection (AR) coated ultra-high-molecular-weight polyethylene (UHMW PE).  The lenses are AR coated high-density polyethylene (HDPE).}
\label{fig:optics_schematic}
\end{figure}

\begin{table}
\caption{The specifications of QUaD.}  \label{tab:spec}
    \centering
    \begin{tabular}{lcc}
      \tableline
       \multicolumn{3}{l}{\em Telescope}\\
        Primary mirror diameter (m)                   & \multicolumn{2}{c}{2.64}\\
        Secondary mirror diameter (m)                 & \multicolumn{2}{c}{0.45}\\
        Total field of view (deg)                     & \multicolumn{2}{c}{1.5} \\
        Pointing accuracy (arcmin)                  & \multicolumn{2}{c}{0.5}\\
        Nominal Frequency bands (GHz)                 & 100 & 150 \\
        Beam FWHM (arcmin)                          & 5.0 & 3.5 \\
        \\
        \multicolumn{3}{l}{\em Receiver}\\
        Band center (GHz)               & 94.6 & 149.5 \\
        Bandwidth (GHz)                 & 26   & 41 \\
        Number of detectors             & 24  & 38 \\
        Number of pixels                & 12  & 19 \\
        Operating pixels (06,07)        & 9   & 18 \\
        Optical Efficiency (\%)         & 27  & 34 \\
        Time constant (ms)              & 30  & 30 \\
        Cross-polar leakage (\%)        & 8   & 8  \\
     \tableline
    \end{tabular}
    \tablecomments{Parameters listed here are average values for the band.  Details are provided in Section \ref{sec:performance}.}
    \vspace{0.125in}
 \end{table}
The QUaD telescope is an axisymmetric Cassegrain design.
The warm foreoptics comprise a parabolic primary mirror and hyperbolic secondary.
The upward-looking cryogenic receiver contains two cooled re-imaging lenses, a
cold stop at the image of the primary, and a curved focal plane.
Figure~\ref{fig:optics_schematic} details the receiver optics chain.
The design requirements were for high image quality (Strehl ratios > 0.98) over a large ($1.5^\circ$) field of view, and a cold stop for sidelobe
control.
Minimizing the secondary blockage required curving the focal plane
and moving the field lens above the primary.

The optical design was primarily performed with the ZEMAX ray tracing software.
Gaussian beam mode analysis and the GRASP8 physical optics package were
used for final optimization, and to investigate the effects of diffraction.
The optical design process is described in greater detail in~\cite{cahill2004} and
\citet{osullivan2008}.

\subsection{Telescope Optics}\label{sec:telescope_optics}
The telescope design, including the use of a foam cone to support the secondary mirror, is closely based on the COMPASS experiment~\citep{farese2004}.
The 2.64\,m, F/0.5 primary mirror\footnote{Costruzioni Ottico-Meccaniche MARCON, Italy} is molded from a single plate of aluminum
with support ribs welded on the rear surface and is identical to the one used in COMPASS.
To minimize movement of the foam cone when the telescope elevation is changed, the secondary mirror assembly was
made as light as possible. For this reason the 0.45\,m secondary mirror was
manufactured from a thin sheet of aluminum supported by a carbon fiber backing.\footnote{Forestal SRL, 248 Via Di Salone, Rome, Italy}

Spillover and sidelobes are a concern for any on-axis telescope design.
Several steps, outlined here, were taken to mitigate their effects.
The receiver optics (see Section~\ref{sec:feeds}) illuminate the primary with a
Gaussian pattern and a -20\,dB edge taper.
A thin aluminium guard ring was used to extend the primary radius by 0.3\,m to
reflect spillover onto cold sky.
The secondary mirror has a 48\,mm hole in its center to prevent re-illumination of the receiver window by the
central portion of the beams from the receiver.
The hole is also used to periodically inject a calibration signal into the data stream from
a source located behind the secondary mirror.
Additionally, an annular aluminum reflecting collar surrounds the window of the cryostat,
filling the geometric shadow of the secondary.
Finally, the telescope is located within a reflective ground shield that blocks
line-of-sight contact between the primary mirror and the ground.
To prevent excess loading from snow accumulation, the ground shield and foam cone were inspected
and swept daily, and detailed logs of snow buildup were kept.
Section~\ref{sec:sidelobes} describes measurements of the sidelobes and their effects.

\subsection{The Foam Cone}
\label{sec:foamcone} The primary reason for using a foam cone to support the secondary
mirror was to preserve the axial symmetry of the instrument which
would have been broken by discrete supporting legs.
However, the cone conferred an extra advantage by trapping warm air venting up from the heated room below
in the space between the cone and the mirror.
This warm air kept the components inside the cone (the receiver snout, secondary assembly, and calibration source) at $\sim 15^\circ$C,
even during the extreme cold of the polar winter.
This protected both the primary and secondary mirrors, and the cryostat
window, from icing and contraction issues and removed concerns
about the cryostat O-rings freezing.

After testing for mm-wave transmission, mechanical stiffness, and
weather resistance, Zotefoam PPA30 (a closed cell propazote foam
expanded with dry nitrogen gas) was selected as the most suitable
material for the cone. The cone was constructed from two layers of $\sim 1.1$"
Zotefoam bonded with Scotch 924 film transfer adhesive. Each
layer is composed of nine sections that were cut from
flat sheets and molded into the appropriate shape by pressing them
between machined aluminum forms while baking them in a custom oven.
The cone assembly was performed on a full-size wooden mandrel.
The two layers of foam are rotated so that the sector seams are anti-aligned.
After assembly the cone and mandrel were placed in a large
plastic bag which was then evacuated to apply uniform pressure
simultaneously over the full cone surface, producing excellent adhesion.
A fiberglass mounting collar clamps the cone along the bottom edge and
provides an attachment point where it is bolted to the primary mirror
guard ring. A similar clamping collar along the top edge forms a mount
for the secondary mirror assembly.

As manufactured, Zotefoam sheets have a $\sim 0.1$" thick over-dense skin on both surfaces.
For improved transmission, the QUaD cone was assembled from sheets that
had the skin removed.
Reflection off a sample section of cone (including two Zotefoam layers and adhesive)
was measured to be 2\% at 150\,GHz.
This reflection coefficient, though small, creates a narrow annular sidelobe at
$100^\circ$ from the telescope boresight (Section \ref{sec:sidelobes}).

\begin{figure}
    \begin{center}
\includegraphics[width=3.3in,clip]{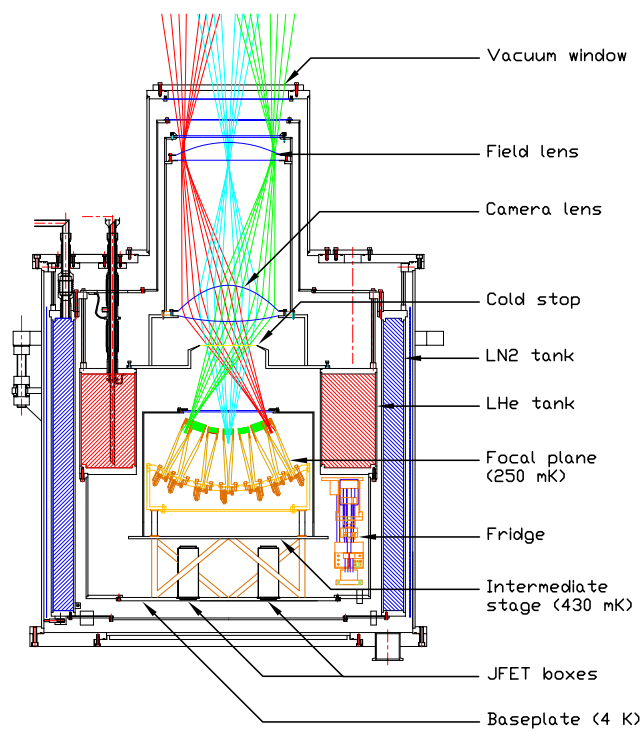}
    \end{center}
    \caption{Schematic of the cryogenic receiver.  Rays are shown for the central and two outer pixels.}
\label{fig:receiver_schematic}
\end{figure}

\subsection{Cryogenic Optics} \label{sec:cryo_optics}
The cryogenic optics couple the Cassegrain focus of the telescope
onto the detectors and form an image of the primary mirror at the cold stop.
Figure~\ref{fig:receiver_schematic} shows the optics mounted in the receiver.
The two lenses are 18\,cm in diameter and are made from high-density polyethylene (HDPE\footnote{Professional Plastics, www.professionalplastics.com}).
The relatively low refractive index of HDPE is desirable to limit losses due to
surface reflection.
The room temperature index of a sample of the HDPE used for the QUaD lenses was
measured using a Fourier Transform Spectrometer (FTS) to be $n=1.5413
\pm 0.0007$ at 150\,GHz.
Using published data on the temperature dependence of the refractive index of HDPE~\citep{birch1984} this was
scaled to 1.583 at 4\,K.
A consequence of the low refractive index of HDPE is that the lenses
are highly curved.

The lenses are anti-reflection coated with a thin film of porous PTFE, resulting in negligible loss,
and are cooled to minimize loading on the detectors.
The location of the field lens complicates cooling, requiring a snout on the cryostat
that protrudes through the hole in the primary mirror.
The lenses are mounted co-axially in an OFHC copper tube that is thermally anchored to the top of the Helium tank.
The temperature of the top lens is 6.2\,K after a cryogen fill and increases by less than 1\,K over
the course of an 18-hour observing run as the liquid helium level decreases.

Corrugated feeds couple the optical signal from the lenses onto the
polarization-sensitive bolometric detectors.
These focal-plane optical components are described in detail in the next section.
A 4\,K knife-edged cold stop, located at a pupil between the camera lens and the focal plane,
truncates the feed horn beams at approximately -20\,dB to prevent the sidelobes from viewing warm
elements further down the optical chain.
The underside of the cold stop is coated in a thin layer of carbon-loaded Stycast for increased
absorption \citep{1994PhDT........23B}.

\section{The Focal Plane} \label{sec:receiver}
\begin{figure}[t]
   \begin{center}
\includegraphics[width=3.25in,clip]{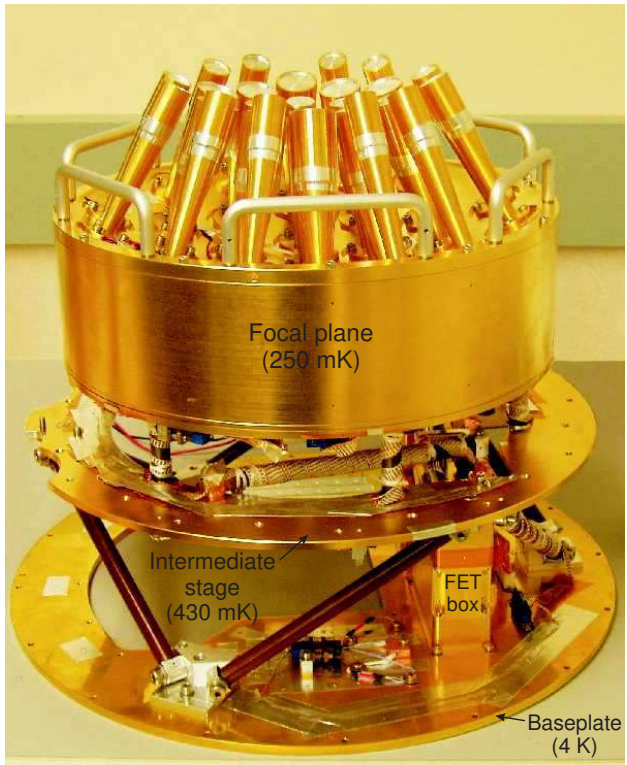}
   \end{center}
\caption{QUaD receiver core including the focal plane, JFET boxes, hexapod, and cold wiring.
The feed horns and cylindrical light-tight bolometer enclosure are at 250~mK. Six Vespel legs
separate this stage from the intermediate temperature ($430$~mK) stage below. A Vespel hexapod
isolates the intermediate stage from the 4~K baseplate.  Low thermal conductivity ribbon
cable, wrapped around the Vespel legs, connects the focal plane to the JFET boxes and
reads out thermometers located on both sub-Kelvin temperature stages. The cutout in the
4~K baseplate is for access to attach the focal plane to the fridge during installation.
Note, in this photograph, optical filters had yet to be installed on the feed horns, which
are shown blanked off with Eccosorb-backed aluminum disks for dark testing. \label{fig:quad_core}}
\end{figure}
\begin{figure}[t]
   \begin{center}
\includegraphics[width=3.25in,clip]{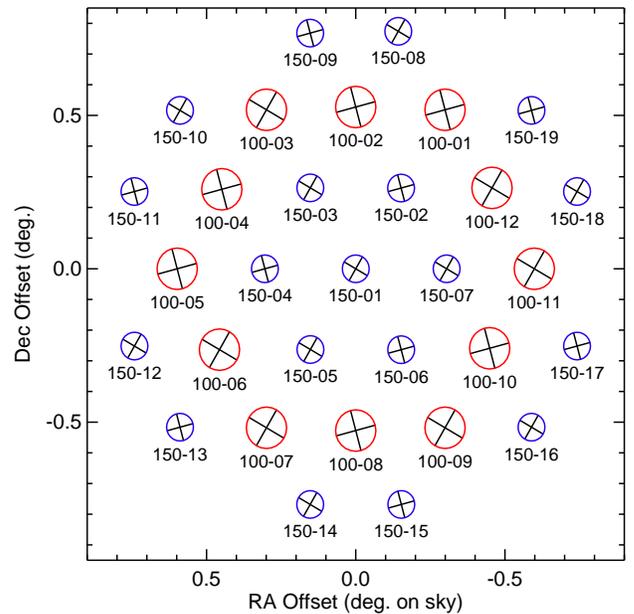}
   \end{center}
\caption{The layout of the QUaD PSBs, showing the relative direction of polarization to which each pixel is sensitive. The 150\,GHz pixels are in blue, the 100\,GHz in red.}\label{fig:fp_layout}
\end{figure}
The QUaD focal plane (Figure~\ref{fig:quad_core}) is a dual-frequency array of 31 pixels, each
composed of a corrugated feed horn and a pair of orthogonally-oriented, polarization-sensitive bolometers.
The focal plane assembly is cooled to $\sim 250$\,mK with a 3-stage sorption fridge and is supported
from the 4\,K baseplate by a two-tiered structure made from low thermal conductivity plastic (Section~\ref{sec:cryo}).

The feed horns are arranged in a hexagonal configuration (Figure~\ref{fig:fp_layout}) and
are positioned so that their phase centers lie along the spherical focal surface
created by the optics.
Nineteen of the feeds operate at 150\,GHz and the remaining twelve operate at 100~GHz.
The focal plane bowl (radius of 175\,mm), which supports the feeds, was milled from a single piece of aluminium 6061 and gold plated to improve thermal conductivity.

Each feed terminates in a pair of orthogonal PSBs that detect the incident radiation.
Summing the two PSB voltages from a pair results in a signal proportional to
the total intensity.
Subtracting the two voltages produces a measurement that is a linear combination of Stokes {\em Q}
and {\em U}, where the relative proportion depends on the orientation angle of the PSB with respect to the sky (Equation~\ref{eq_v_psb}).
In order to determine the values of both {\em Q} and {\em U} for a given spot on the sky,
it must be observed at two different orientation angles.
For QUaD, the voltage from each PSB is independently recorded.
Summing and differencing is performed in software during post-processing.

Maps of the sky are made by scanning the telescope in azimuth with the
detector rows aligned along the scan direction.
The pixel orientations were chosen so that a given location on the
sky is observed at two different angles by the detectors in a given row during each scan.
Additionally, two different values of the telescope ``deck'' rotation angle ($0^\circ$ and $60^\circ$) were used while mapping
to better constrain the polarization angles of the source and as a check for systematic effects.

A cylindrical light shield, also made from aluminium, surrounds the underside of the focal plane bowl
and encloses the detectors, load resistor boards (Section~\ref{sec:cold_electronics}) and miscellaneous
thermometry.
Two thermistors and three heater resistors are mounted on the focal plane bowl and
are used with an external control system to stabilize the focal plane temperature during operation (Section~\ref{sec:thermometry}).
Four ``dark'' PSB modules are mounted to the back of the focal plane and are used to monitor for non-optically induced
contamination due to, for example, temperature drifts or electrical pickup.
Several fixed resistors are also included and are used to monitor the noise in the readout electronics.

\subsection{Spectral Bands and Filtering}\label{sec:filters}
\begin{figure}[t]
   \begin{center}
\includegraphics[width=3.35in,clip]{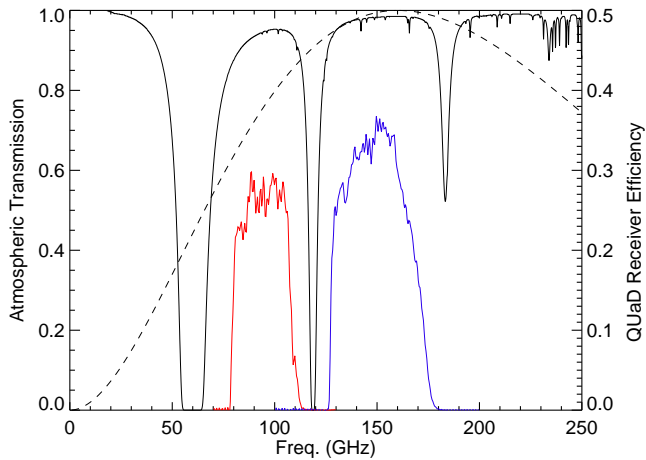}
   \end{center}
\caption{QUaD average spectral bands (red, blue), the South Pole
atmospheric transmission (solid black), and the CMB spectrum (dashed black). The QUaD bands
are normalized in terms of absolute transmission per polarization referring to the scale on the right.}\label{fig_avg_spectral_bands}
\end{figure}
The two QUaD bands of 78-106 and 126-170 GHz were
chosen to span atmospheric windows of high transparency (Figure
\ref{fig_avg_spectral_bands}).  They are near the maximum of the 2.7~K CMB
blackbody spectrum and the minimum of polarized foreground
contamination from galactic dust and synchrotron
emission~\citep{2007ApJ...665..355K}. Observing at two frequencies
allows the level of foreground contamination in the final maps to be
investigated.

The band edges are set entirely using optical methods. The waveguide
cutoff in the narrow throat section of each feed horn sets the lower
band edge (Section~\ref{sec:feeds}) while metal-mesh low-pass
filters on the horn apertures set the upper band edge. The low-pass
filters were manufactured at Cardiff University using
photolithography to pattern capacitative structures on thin layers of
vacuum-deposited metal over a polypropylene substrate \citep{ade2006}.
Multiple layers with staggered cutoff frequencies are required to block
leaks that occur at the harmonics and give the required rejection of out-of-band power.
The filters are anti-reflection coated to improve in-band transmission.

The particular filter combinations at each frequency
(Figure~\ref{fig:corrugated_feeds}) were chosen after extensive testing
of different filters in a single-pixel optical testbed.
An FTS was used to measure the band shape and thick grill filters were used to check for out-of-band
leaks.\footnote{Thick-grill filters are metal plates with drilled holes
that act as high-pass filters via their waveguide cutoff.  Filters with
holes sized to cut on just above the band of interest allows the
above-band power to be measured.} FTS measurements of all feeds were
made prior to shipping the receiver to the Pole and again before the
receiver was installed on the telescope. By adding or removing filters
from the chain during laboratory testing we were able to verify that
the filters contribute negligible cross-polar leakage (Section~\ref{sec:pol}).

Blocking filters that reflect high-frequency out-of-band radiation are located on
each thermal stage within the cryostat, reducing the radiative load on the colder stages.
They are held in place with aluminium clamping rings that are screwed down for good
thermal contact.
The filters at the horn apertures are secured with a thin-walled brass filter
cap that slides over the horn body and is itself held in place with
aluminium tape.  A beryllium copper wavy spring washer provides thermal
contact between the filter stack and the horn mouth, and a thin ring of
indium wire seals the filters against the cap to prevent high-frequency
optical leaks.

\subsection{Feed Horns}
\begin{figure}[t]
   \begin{center}
\includegraphics[width=2.75in,clip]{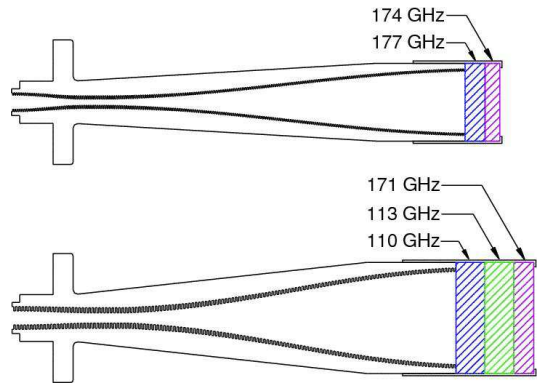}
   \end{center}
\caption{Schematic of the QUaD corrugated feeds, band-defining filters, and filter caps. The length of the
100~GHz (150~GHz) feed is 100~mm (102~mm), not including the filters and filter
cap.\label{fig:corrugated_feeds}}
\end{figure}
\begin{figure}
    \centering
\includegraphics[width=3in,clip]{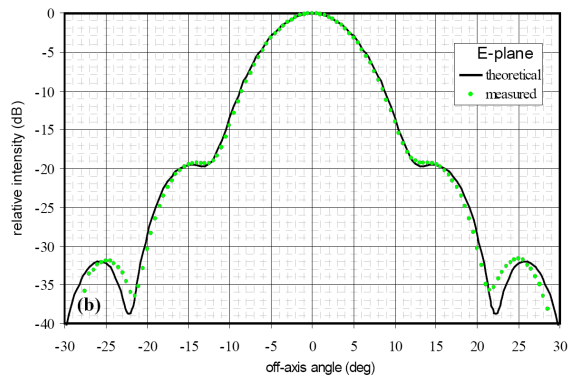}
    \caption{The measured and predicted beam pattern from a QUaD 100 GHz corrugated feed horn.}
    \label{fig:feed_beam}
\end{figure}
\label{sec:feeds} Corrugated feed horns couple the telescope optics to the detectors.
The $\lambda/4$ corrugations allow the hybrid $\mbox{HE}_{11}$ mode, which is a linear
combination of the $\mbox{TE}_{11}$ and $\mbox{TM}_{11}$ circular guide
modes, to propagate. This mode possesses a nearly axially-symmetric
field amplitude that smoothly tapers to zero along the horn walls,
resulting in nearly symmetric E and H plane beam patterns, low
cross-polar response, and low sidelobes.  Consequently this type of
horn is ideal for millimeter-wave polarimetry.

Thomas Keating Ltd.\footnote{Station Mills, Billingshurst, West Sussex, RH14 9SH UK}
manufactured the horns by electroforming copper onto CNC-machined
aluminum mandrels, which were subsequently dissolved.
Figure \ref{fig:corrugated_feeds} shows a cross-sectional schematic of
the QUaD horns. The feeds were designed to be roughly matched in length
at each frequency to prevent shadowing effects on the focal plane. Note
that QUaD employs ``profiled'' feeds in which the diameter expands
non-linearly along the length of the horn. Profiling allows the horns
to be shorter than a conical feed with equivalent sidelobe
levels~\citep{gleeson2005}. This is particularly important for QUaD
because the curved focal surface causes both the height and diameter of
the required cold volume in the receiver to depend on the feed length.

The horn beams were modeled using a waveguide mode matching technique
and the resulting predictions were checked against measurements of
prototype horns (Figure \ref{fig:feed_beam}). The horns were designed
to illuminate the cold stop with an edge taper of approximately -20~dB;
however when the filter cap, the filters, and the effects of the broad bandwidth are
included, the feed beams are slightly wider.
See \citet{murphy2005} for further detail on
the design, numerical modeling, and experimental verification of the
QUaD corrugated feeds.

The narrow waveguide throat acts as a high pass filter, defining
the lower edge of the frequency band (Section~\ref{sec:filters}).
\citet{gleeson2005} provides more detail on design
considerations for this section of the feed.  After the throat, the
back end of the feed flares outward to couple the field
onto the PSB absorber (Section~\ref{sec:psbs}). Modeling of the feed
indicates a
peak cross-polar level $\sim 30$~dB down from the co-polar beam, well below
the levels introduced by the detectors.

\subsection{Polarization-Sensitive Bolometers}
\begin{figure}[t]
   \begin{center}
\includegraphics[width=3.0in]{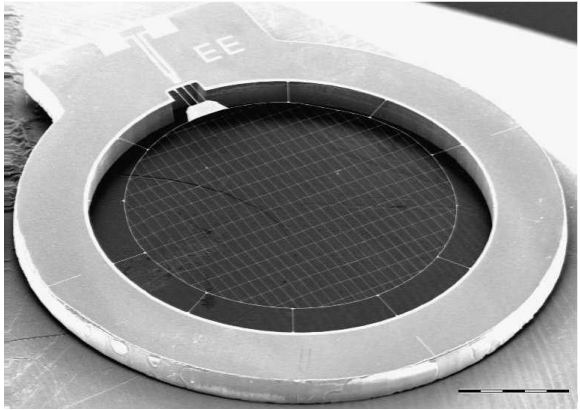}
   \end{center}
\caption{Scanning electron micrograph of a single PSB membrane.  The 4.5\,mm diameter absorber is suspended with an array of radial supports.  Metalized traces (running from the upper left to the lower right) selectively absorb one linear polarization of incident optical radiation.  Perpendicular, non-metalized traces provide mechanical support.  The 150\,$\mu$m spacing of the conductors allows the same membranes to be used for the 100 and 150\,GHz bands.  A thermistor, located on the edge of the absorber (upper left), measures the temperature.  Three thicker legs (upper left) set the thermal conductance to the housing (known as $G$).  The center leg also caries the electrical leads from the thermistor to a pair of wirebond pads.  The outer two legs could be trimmed, allowing $G$ to be tuned to one of four possible values.
\label{fig:PSB_pictures}}
\end{figure}
\label{sec:psbs}
Bolometers detect incident optical power by measuring the temperature increase of
an absorber that is weakly thermally connected to a fixed bath temperature.
QUaD uses polarization-sensitive bolometers (PSB) where the absorber is patterned
as an array of parallel line conductors, so that it only couples to one linear polarization (Figure~\ref{fig:PSB_pictures}).
These detectors were fabricated at the JPL Micro Devices Laboratory, where
similar detectors were developed for the Planck
satellite \citep{2003SPIE.4855..227J} and the Boomeramg experiment \citep{masi2006}.
PSBs identical to those in QUaD are used by the BICEP
experiment \citep{yoon2006}.

The PSB membranes are fabricated from silicon-nitride (Si$_3$N$_4$) and metalized with 120\,\AA~of gold over 20\,\AA~of titanium.
The width of the conductors and their metalization thickness were
optimized for maximum coupling using numerical simulation and experimentation with trial devices at Caltech/JPL.
The conductor spacing of 150\,$\mu$m allows the same membranes to be used at both observing frequencies.
Non-metalized legs run perpendicular to the conductors for mechanical support and have
negligible impact on the optical properties \citep{2003SPIE.4855..227J}.

An array of non-metalized radial spokes support the circular absorber.
A neutron-transmutation-doped (NTD) Ge thermistor\footnote{J. Beeman, www.haller-beeman.com}
monitors its temperature.
The thermistor is located on a conducting ring that runs along the circumference of the absorber.
The ring lies outside the optical coupling area and provides a thermal connection between the parallel conducting traces of the absorber.

The dominant thermal path to the absorber is through three thicker legs near the thermistor,
the central of which also carries the electric signals.
These legs were designed to be laser ablated, if necessary,
allowing the overall thermal conductance between the absorber and the
focal plane, known as $G$, to be tuned.
This operation was not performed for QUaD.

For ease of handling and attachment to the feed horns, pairs of orthogonally-oriented PSBs
are mounted in protective brass modules.
The module forms a cylindrical integrating cavity, $\lambda / 2$ in length, so that
a standing wave is created and the electric field is maximized in the center where the PSBs are located.
The two absorbers are coaxial and are separated by only 100 $\mu$m to ensure that they
sample the same electric field.

The PSB modules are referenced in position and heatsunk to the back flange of the
feedhorns.
The angular alignment of each module is set by a stainless steel dowel pin on the module and
a machined radial slot in the back of the focal plane.
The dominant angular alignment error is in the placement of the silicon nitride absorbing grids within the modules, which is done under a microscope.
Section~\ref{sec:loadcurves} reviews the characterization of the detectors, and
Section~\ref{sec:pol} discusses the measurements of their polarization properties.

\section{Cryogenics and Thermometry} \label{sec:cryo}
\subsection{Cryostat}  \label{sec:cryostat}
The QUaD receiver is housed in a custom-made liquid helium/liquid nitrogen cryostat (shown in schematic in Figure~\ref{fig:receiver_schematic})
manufactured by AS Scientific.\footnote{Abingdon, UK.
www.asscientific.co.uk}  The instrument is upward looking, with the
cold re-imaging optics housed in a snout that is inserted through the
hole in the center of the primary mirror.
The receiver vacuum window, is located on top of the snout.
It is made from ultra high molecular weight polyethylene, roughly 1\,cm thick, and
surfaces are anti-reflection coated to improve transmission.

The focal plane and cryogenic electronics (Sections~\ref{sec:receiver} and \ref{sec:cold_electronics}) are assembled on a removable central core that is enclosed by concentric, toroidal liquid cryogen tanks.
The capacities of the tanks are 35\,L for liquid nitrogen and 21\,L for liquid helium, resulting in hold times of 3.5 and 1.3 days respectively, including the daily fridge cycle.
During normal observation, both tanks are filled once per day.

A G-10 fiberglass truss frame supports the tanks, providing rigidity with low thermal conductivity. Sections of thin-walled stainless steel bellows limit the conductivity through the fill tubes and allow for differential contraction of the different thermal stages. The cryostat has four fill tubes (fill and vent per tank), two of which house gauges to monitor cryogen levels. Because the fill tubes are inaccessible when the cryostat is installed on the telescope, refills are performed via flexible transfer lines, about 3.5\,m long, made out of two sections connected through standard bayonet fittings. The shorter section is inserted in the cryostat fill tube and remains in place throughout the observing season.  The longer section is only connected during refills.

Within the cryostat, both liquid cryogen tanks, as well as the aluminum radiation shields thermally connected to them, are wrapped in multi-layer aluminized mylar insulation.
Filtered apertures on the top of each shield allow the optical signal to reach the focal plane, while blocking out-of-band IR radiation (see Figure~\ref{fig:optics_schematic}).
The focal plane is surrounded by an additional radiation shield that is thermally connected to the intermediate stage of the sub-Kelvin refrigerator, maintaining a temperature of 430\,mK.
The large focal plane mass ($\sim 10$\,kg) combined with the thermally isolating support structure results in a long cool down period. Starting from room temperature, both tanks are initially filled with liquid nitrogen. Approximately five days later, the focal plane reaches 100\,K at which point the inner tank is filled with liquid helium. Two more days of pre-cooling are then required before the fridge can be cycled.

In order to speed up the cool down, two active He-3 gas-gap heat switches are used.\footnote{Chase Research Cryogenics, 140 Manchester Rd, Sheffield S10 5DL (UK)}
The first connects the fridge 4\,K baseplate to the intermediate stage.
The second connects the fridge intermediate stage to the ultracold stage.
These switches are turned off once the focal plane temperature approaches 4\,K.
The heat switch connected to the fridge ultracold stage has an estimated off-conductance $< 0.1\,\mu$W.

\begin{figure}[t]
\begin{center}
\includegraphics[width=3in,clip]{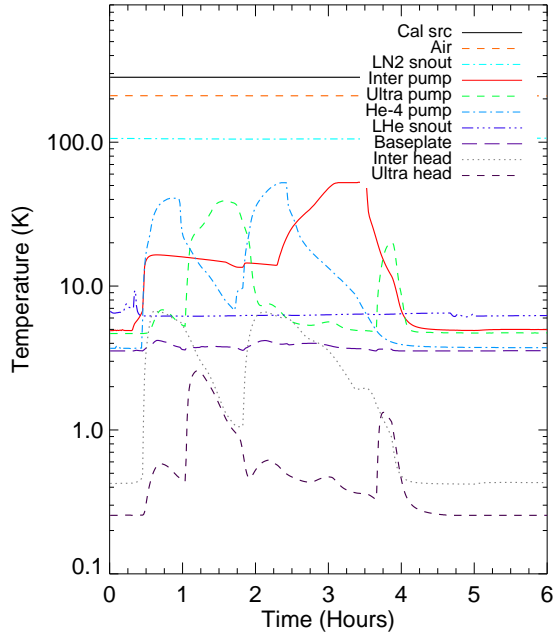}
\end{center}
\caption{Temperatures of QUaD components through a fridge cycle at the start of an observing day.  ``Cal src'' is the temperature of the calibration source located above the secondary mirror, inside the foam cone (Section~\ref{sec:calibration}).  ``Air'' is the external air temperature.  ``LN2 snout'' and ``LHe snout'' are the temperatures at the top of the liquid nitrogen and liquid helium cryostat snouts.  The remaining items are the temperatures of the fridge pumps, cold heads, and baseplate (Section~\ref{sec:fridge}).} \label{fig:temperatures}
\end{figure}

\subsection{Fridge}\label{sec:fridge}
The focal plane is cooled to an operating temperature of $\sim$250\,mK
using a 3-stage He-4/He-3/He-3 sorption refrigerator, manufactured by Chase Research Cryogenics \citep{bhatia2000}. Similar fridges have been used successfully in bolometer-based receivers such as ACBAR and
BOLOCAM \citep{runyan2003,glenn1998}.
The first and second fridge stages, known together as the Intercooler
(IC), contain, respectively, He-4 and He-3. The third stage contains He-3 and
is referred to as the Ultracooler (UC).
The fridge is cycled by first condensing He-4 and using the enthalpy of that liquid to cool the IC and UC
condensation points below the He-3 critical temperature.
During normal operations at the South Pole, the fridge IC and UC stages achieve
temperatures of 430\,mK and 253\,mK.
The heat loads on the fridge stages are measured to be $\sim 40$\,$\mu$W on the IC and $< 0.5\,\mu$W on the UC.

To achieve better condensation efficiency during the cycle, the fridge is mounted directly on the LHe tank, in a position that keeps the fridge baseplate in contact with liquid helium when the telescope is at the typical observing elevation of $\sim 50^\circ$.
To preserve this contact, rotation of the system about the optical axis (``deck'' rotation) is limited to $\sim \pm 60^\circ$ during normal observing.

The large focal plane mass ($\sim 10$\,kg) presented a particular
challenge to achieving an optimized fridge cycle.
A maximum possible hold time and minimum loss of observing due to the cycling time were desired.
The procedure that was developed (shown in Figure \ref{fig:temperatures})
allowed the QUaD fridge to be cycled in approximately four hours and results in
hold times for the UC and IC stages of 24 and 31 hours, respectively
(duty cicle $\sim 83$\%).  Once the optimized cycle was determined, the
procedure was automated (Section \ref{sec:thermometry}) so that the operator can perform
a fridge cycle by issuing a single command on the control system.

\subsection{Receiver Core} \label{sec:sciencecore}
The receiver core comprises the focal plane and associated mechanical
and electrical support hardware (Figure~\ref{fig:quad_core}).
It is composed of three stages that are each held at a different temperature during operation:
a 4\,K baseplate, a 430\,mK intermediate stage, and the 250\,mK focal plane assembly.

The 4\,K baseplate is made of gold-plated 6061 aluminum.
It supports the rest of the focal plane structure and provides the mounting point for the two JFET amplifier boxes (Section~\ref{sec:electronics}).
When installed in the cryostat, the baseplate is mechanically attached to the liquid helium tank through a cylindrical aluminum radiation shield.
The dominant thermal path is through two parallel OFHC copper straps that connect to the helium tank.
This insures that the baseplate stays near 4\,K despite the 26\,mW dissipation from the JFETs.

The intermediate stage is a gold-plated aluminum ring that is thermally connected to the fridge IC.
It is mechanically attached to the 4\,K baseplate by a six-legged hexapod structure of 6" SP-1 Vespel tubes (0.438" outside diameter and 0.031" wall thickness).
The focal plane assembly is mounted to the intermediate stage with
six shorter legs made from SP-22 Vespel\footnote{http://www2.dupont.com/Vespel/en\_US/} tubing machined to the same diameter and wall thickness.
The focal plane is thermally connected to the fridge UC.

The estimated heat load on the fridge from the Vespel supports is
0.13\,$\mu$K on the UC and 24\,$\mu$K on the IC.
SP-22 is used for the focal plane supports because it has lower thermal
conductivity than SP-1 in the sub-Kelvin temperature range \citep{marcus_thesis}.
Both the focal plane and the intermediate stage are attached to their
respective fridge cold heads via flexible, copper heat straps made from
braided OFHC copper electrical shielding.
Multiple braids are twisted together to increase the cross-sectional area
and the straps are annealed for increased conductivity.

Because of the high electrical impedance of the bolometers, microphonic pickup
from vibrations of the wiring is a concern (Section
\ref{sec:interference}). For this reason all of the wiring between the
JFET amplifiers and the bolometers must be rigidly supported. The
Vespel legs that support the focal plane provide natural attachment
points for the wiring that runs between the stages (Figure
\ref{fig:quad_core}). All of the non-isothermal wires in the cryostat
are manganin (0.003") ribbon cables woven into a robust ribbon cable
with nylon thread for strain relief.\footnote{www.tekdata.co.uk} The
wiring is tightly wrapped around the Vespel legs along a helical path
and fixed down at regular intervals using a combination of teflon tape
and lacing tape.
Two additional Vespel tubes between the intermediate stage and the JFET modules act as a bridge to support the wiring
along this critical signal path.
The heat load on the fridge UC stage is reduced by heat sinking the focal plane wiring to the intermediate temperature stage.
The heat load from the wiring is estimated to be $\sim 8$\,$\mu$W on the fridge IC and $< 0.1\,\mu$W on the UC.
On the isothermal stages, including the back of the focal plane bowl, stiffer (28
AWG) copper wiring is used and is held down by aluminium tape.
Specially-designed aluminum brackets support the wiring at all connector interfaces, including the PSB modules.

\subsection{Thermometry and Temperature Control} \label{sec:thermometry}
The temperatures of all the major cryogenic components are monitored,
including the focal plane, the fridge pumps, cold heads and
baseplate, the snout that holds the two lenses, and the liquid cryogen
tanks. Silicon diode sensors\footnote{Lake Shore DT-470 series, Lake
Shore Cryotronics, Inc. www.lakeshore.com} monitor the
temperatures of components that operate at or above 4\,K. Germanium
resistance thermometers\footnote{Lake Shore GR-200 series} (GRTs),
monitor the temperatures of sub-Kelvin components. The diodes that are
mounted on the focal plane allow the monitoring of the cool down from
room temperature; however, their large power dissipation ($\sim
15\;\mu$W) requires that they be switched off during sub-Kelvin
operation.

Figure \ref{fig:temperatures} shows temperature readouts during and shortly after a routine fridge cycle.
In order to monitor the temperatures of all the different cryogenic components and to operate the heaters of fridge pumps and gas-gap heat switches, required to cycle the sorption fridge, we use custom-made thermometer readout and control electronics. With a very compact design (only one main unit and one separate power supply unit), it can readout up to 6 GRTs and 21 diodes, and operate 5 heaters. An embedded computer, the TINI\footnote{MAXIM,www.maxim-ic.com}(Tiny INternet Interface), reads the digitized voltage outputs, sets the heaters drives, and communicates
with the control system.
With this system it is possible to cycle the fridge with no external intervention, by running a script on the TINI.
The same electronics drives two additional heaters (with constant drive and manual control) for the gas-gap heat switches we use to pre-cool fridge and focal plane at liquid helium temperature.

The focal plane temperature is stabilized with a separate system.
Temperature is read out with a custom AC bridge connected to an NTD Ge thermistor (Haller-Beeman)
located on the focal plane.
The bridge operates at the bolometer bias frequency.
A Stanford Research Systems SIM960 PID controller, servoed off the
thermistor readout, drives three heater resistors in parallel, located symmetrically around the joint between the fridge heat strap
and the focal plane.
An additional thermistor, read out with the same system, is used as a temperature monitor.
During routine observing, the PID set point is 258\,mK, $\sim 5$\,mK above the natural operating temperature
of the fridge, requiring $\sim 0.3$\,$\mu$W of electrical heater power.
This maintains a stable temperature despite the telescope motion associated with raster scanning
and the slight warming of the fridge UC that occurs over the course of an 18 hour observation. 

\section{Electronics} \label{sec:electronics}
 \begin{table}
    \caption{Electronics parameters.}  \label{tab:electronics}

    \begin{center}
    \begin{tabular}{lc}
     \tableline
        Number of channels                    & {96} \\
        DC mode gain                          & {200} \\
        AC mode gain                          & {$10^5$} \\
        Capacitance\tablenotemark{a} (pF)     & {85} \\
        Load resistance (M$\Omega$)           & {40} \\
        Bias generator noise (nV Hz$^{-1/2}$) & {3} \\
        JFET noise (nV Hz$^{-1/2}$)           & {7} \\
        Warm amplifier noise (nV Hz$^{-1/2}$)      & {5} \\
        Total electronics noise (nV Hz$^{-1/2}$)   & {9} \\
        $1/f$ knee (mHz)                      & {10} \\
        AC bias frequency\tn{b} (Hz)          & {110} \\
        AC bias current\tn{c} (nA)            & {1.25} \\
        Readout bandwidth (Hz)                & {20} \\
        Sampling frequency (Hz)               & {100} \\
        Power\tn{d} (W)                       & {60} \\
     \tableline
    \end{tabular}\\
    \end{center}
    
    \footnotesize
    $^{\mbox{\scriptsize a}}$The capacitance between the two signal leads of a channel arising from the connectors and wiring between the PSBs and the JFETs.\\
    $^{\mbox{\scriptsize b}}$The bias frequency is adjustable from 40 to 250 Hz.  The frequency is chosen to minimize interference (Section \ref{sec:interference}).\\
    $^{\mbox{\scriptsize c}}$The bias current is adjustable from 0 to 30 nA.  The current is a tradeoff between stability and sensitivity (Section \ref{sec:performance}).\\
    $^{\mbox{\scriptsize d}}$Power for the amplifier boxes, not including the commercial ADC system.\\
    \vspace{0.05in}
 \end{table}

\begin{figure*}
   \begin{center}
\includegraphics[width=5.5in,clip]{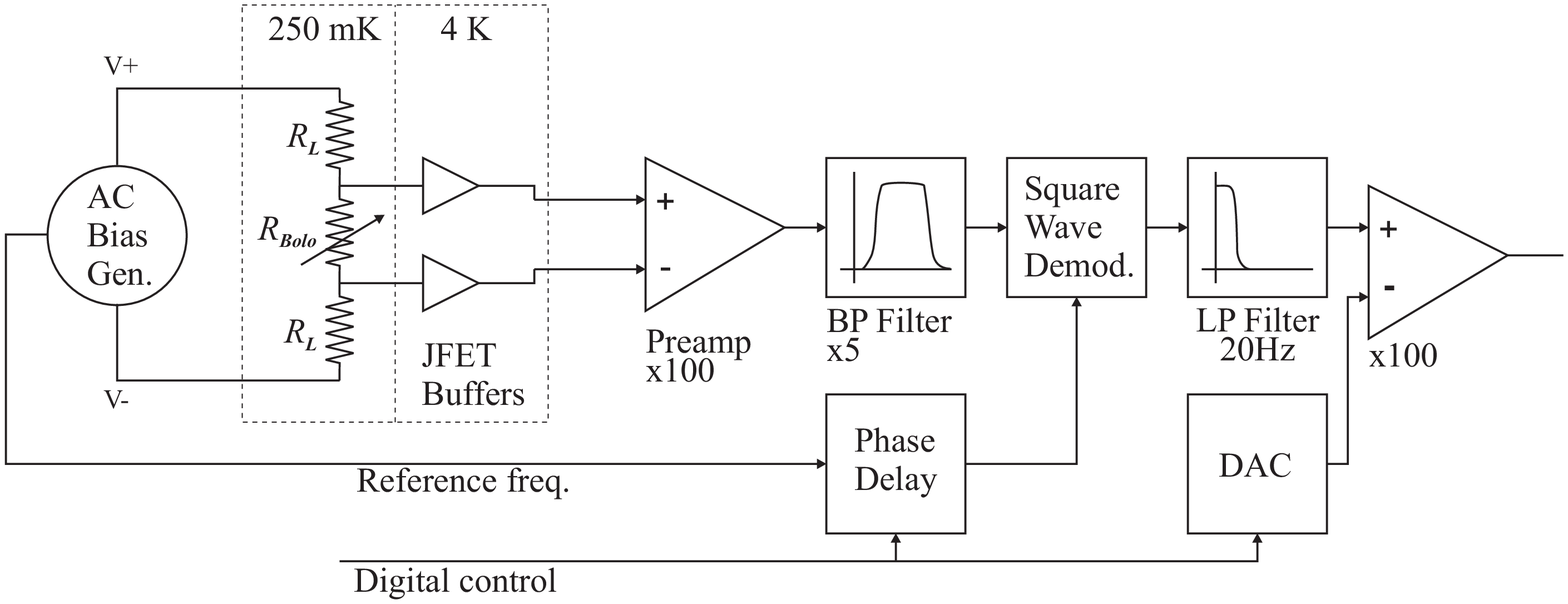}
   \end{center}
   \caption{Block diagram of the QUaD readout electronics.  The components within the dashed lines are located in the cryostat.  The load resistors and detectors are on the focal plane assembly and operate at approximately 250\,mK.  The JFET amplifiers modules are mounted on the 4\,K baseplate; however, within the modules the JFETs themselves operate at roughly 120\,K (Section~\ref{sec:electronics}).} \vspace{0.2in}
   \label{fig:electronics_overview}
\end{figure*}

\label{sec:readout} The QUaD electronics employ an AC-biased, fully
differential readout to amplify the bolometer signals.
The basic scheme has a long history in bolometric CMB instruments
\citep[see][for instance]{holzapfel1997,glenn1998,crill2003}.
An overview of the electronics chain is shown in Figure \ref{fig:electronics_overview}.
The bias generator excites the bolometers with a sinusoidal current
through a pair of equal valued load resistors. The balanced nature of
the bolometer/load resistor bridge minimizes crosstalk and pickup along
the high-impedance wiring leading to the cryogenic JFET buffer
amplifiers. The warm electronics amplify, demodulate, and filter the
output from the JFETs. Finally, the data acquisition system digitizes
and archives the processed signals. The QUaD electronics are interfaced
to the control software to allow the bias frequency and amplitude,
amplifier gain, DC offset removal and phase adjustment to be set
remotely. The electronics development and testing are described in more
detail in \citet{hinderks2005}.

\subsection{Cold Electronics} \label{sec:cold_electronics}
\begin{figure}[t]
   \begin{center}
\includegraphics[width=3.0in,clip]{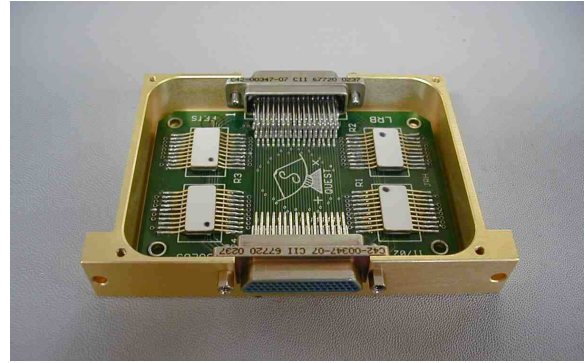}
   \end{center}
\caption{Photograph of a QUaD load resistor board. Each board contains
56 resistors (including eight spares) in
four surface-mount packages and services up to 24 channels. The boards
themselves are with double-sided 1 oz Copper and 8 mil traces. Input
and output from the board is via surface mount 51-way micro-D
connectors (Cristek Interconnects Inc. www.cristek.com). These
connectors have long surface mount leads that take up differential
contraction between the fiberglass board and the aluminum mounting
structure that provides mechanical support.}
   \label{fig:load_resistor_board}
\end{figure}

The load resistors are Nichrome metal film on a silicon substrate in a
surface-mount package.\footnote{Mini Systems, Inc. www.mini-systems.biz}
There are two $20\:\mbox{M}\Omega$ load resistors in each circuit. This
value was chosen to be much larger than the bolometer
resistance to provide a bias current that remains approximately
constant for small loading changes. The load resistor boards
(Figure~\ref{fig:load_resistor_board}) are located on the 250\,mK stage
to reduce Johnson noise. The boards were fabricated from standard 1/16"
FR-4 substrate. Input and output from the board is via surface mount
51-way micro-D connectors.\footnote{Cristek Interconnects Inc.
www.cristek.com} These connectors have leads that were
specifically designed to take up differential contraction between the
fiberglass board and the aluminum mounting structure that is used to
provide mechanical support. In three years of sub-Kelvin operation and
numerous thermal cycles, we did not experience a single failure.

The JFET amplifiers are located in two boxes mounted on the 4\,K
baseplate, directly below the focal plane assembly. The transistors
themselves operate at an elevated temperature of $\sim 120$\,K for
optimum noise performance and consequently need to be thermally
isolated from their 4\,K enclosure. Each box contains two
silicon nitride membranes, each holding 24 dual JFET
dies\footnote{Siliconix U401} allowing readout of 48 channels per box. The
membranes provide thermal isolation of the amplifiers from the 4\,K stage
and are mechanically supported by the micro-D input/output connectors.
The JFETs are configured as source followers with their drains
connected to a common positive supply and the sources connected via
125\,k$\Omega$ resistors to the common negative supply. The power supply
voltages for each membrane is adjusted for lowest
noise performance. Each membrane dissipates $\sim 6.5$\,mW during normal
operation with heater resistors used to assist at startup. The low
thermal mass allows the membranes to reach a stable operating
temperature in approximately one minute.

\subsection{Warm Electronics} \label{sec:warm_electronics}
The bias generator is used to provide a stable sine-wave excitation
that can be adjusted over a frequency range of 40-250\,Hz with 10~bit
resolution.  The sine wave is generated by filtering a square wave with
a Q=10 bandpass filter.  A low-noise DC voltage reference and an
analog modulator switched by a crystal oscillator generate the square
wave. The bandpass filter is made electronically tunable by using multiplying DACs
in place of fixed resistors.
The bias frequency is common to all bolometers but the 100\,GHz and 150\,GHz
bias amplitudes can be independently adjusted. A DC bias mode is
provided for testing and is used for taking bolometer I-V curves
(Section~\ref{sec:loadcurves}). The bias amplitudes are archived along
with the bolometer data. The bias frequency, amplitude and bandpass
tuning are computer controlled via a serial interface.

The bolometer signals are amplified with low-noise AD624 amplifiers
and then filtered with a broad bandpass filter (Q~$\sim 0.5$) that is
designed to have a close to flat response over the available range of
bias frequencies. The signals are then multiplied by a square-wave
reference from the bias board to demodulate the component at the bias
frequency. An adjustable delay circuit for each channel compensates the demodulator
reference signal for phase shifts within the cryostat. A steep low-pass
filter (20 Hz, 6-pole) follows the demodulator.

To ease the dynamic
range requirement on the ADC system, a 12-bit DAC is adjusted
periodically to null out the large DC component of the demodulated
signal. This is followed by an additional gain stage of 100.
During observation, the DC-offset removal DAC settings are adjusted
approximately every thirty minutes to prevent any channels saturating
due to $1/f$-noise or elevation changes.

The electronics can be switched to a ``low-gain'' mode in which the
DC-offset removal and additional gain stage are bypassed.
This mode is useful for testing, and is used routinely by the control
system to automatically determine the appropriate setting for each
channel's DC-offset removal DAC.
Finally, buffer circuitry creates a balanced differential output for
driving the cabling to the ADC. All of the settings including reference
phase delay, DC offset removal, and gain mode are controlled via a
serial interface.

The bias and readout electronics are located in RF boxes attached via
an RF-tight interface box directly to the bottom surface of the
cryostat. All wiring entering the cryostat passes through these
interface boxes and is RF filtered using filtered D-sub connectors.\footnote{Spectrum Control Series 700 with 1000 pF PI filters}

\subsection{Data Acquisition and Control System}
Realtime operations including telescope control, and digitization of
the bolometer data, is handled by a VME controller running VXWorks in a
crate mounted adjacent to the cryostat. Two 64-channel, 16-bit ADC
cards digitize the bolometer data. A 32-channel DIO card provides the
control interface to the bias and readout boards. A Linux-based PC
system provides the user interface and data archiving for the realtime
controller.

Once per second, the software archives a  complete snapshot of the
system, known as a register frame. Each register frame contains the
most recent values from all of the attached hardware, including
thermometry, bias settings, and DC offset removal values. The bolometer
data along with the telescope axis encoders are sampled at 100\,Hz, with
100 samples per channel stored in each register frame.
Software commands are provided for controlling all the components of
the system. These commands are combined to form complete
observing scripts. A client program allows users to run scripts
and monitor the data in realtime.

\section{Instrument Characterization} \label{sec:performance}
\subsection{Telescope Pointing}\label{sec:pointing}
\begin{figure}
    \centering
\includegraphics[width=3.25in]{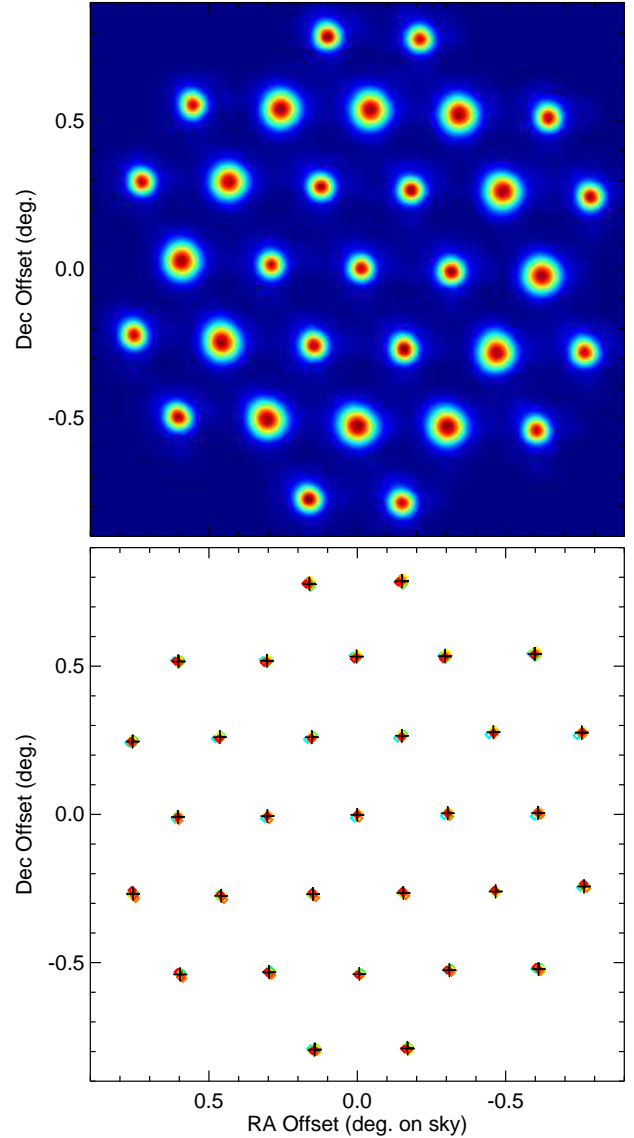}
    \caption{(top) Raster map over the galactic source RCW38.  The map has been smoothed with a $1.2^\prime$ Gaussian. (bottom) Radio pointing locations for each channel derived from twelve full raster maps spread over the second and third seasons.  The black crosses indicate the mean offset for each channel.}\label{fig:fp_plots}
\end{figure}
\begin{figure}
\begin{center}
\includegraphics[width=2.75in,clip]{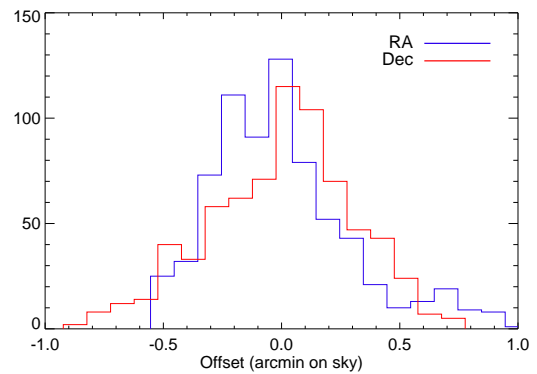}
\end{center}
\caption{Histogram showing the variation in the derived radio pointing for all feeds for seasons two and three, from the data plotted in the bottom panel of Figure \ref{fig:fp_plots}.  The RMS of both distributions is $\sim 0.3^\prime$.
 \label{fig:rcw38_beam_centroid_histograms}}
\end{figure}
\begin{figure}
\begin{center}
\includegraphics[width=3.25in,clip]{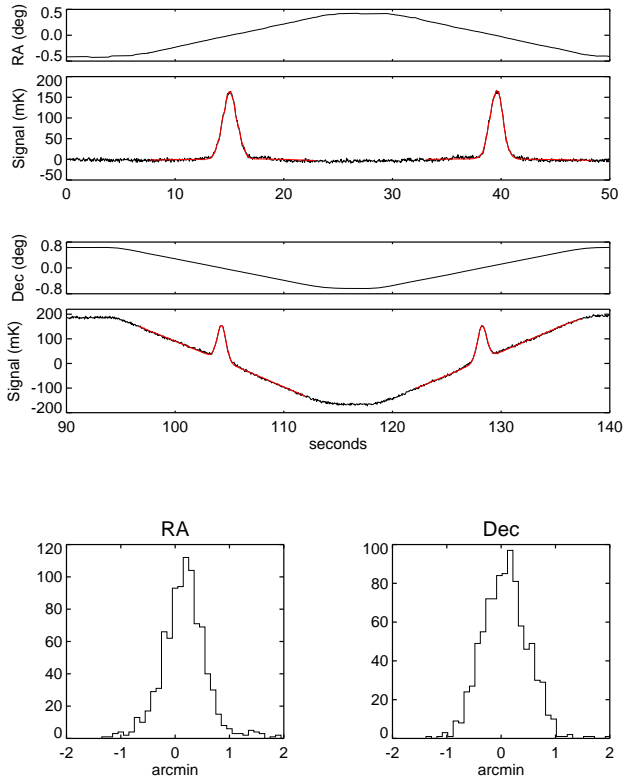}
\end{center}
\caption{The two upper panels show a ``pointing cross'' observation in which the central pixel is scanned across a bright source, first in RA, then in Dec.  The lower panels shows the variation of the derived pointing offset over the course of a season.  The RMS of both distributions is $\sim 0.4^\prime$.\label{fig:cross_obs_reduction.eps}}
\end{figure}
\begin{figure}
\begin{center}
\includegraphics[width=3.4in]{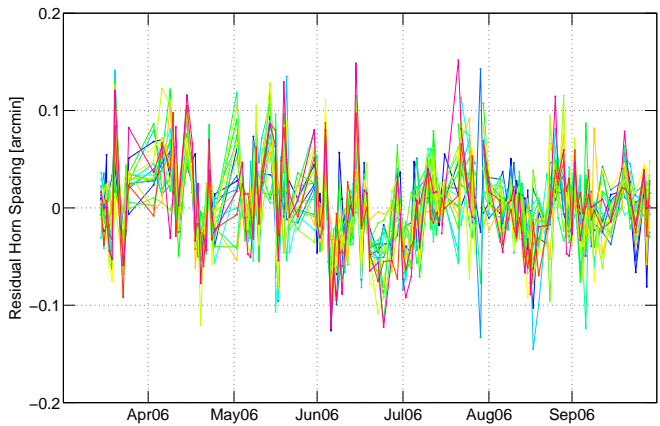}
\end{center}
\caption{Residual separation between pairs of adjacent feeds as measured by daily azimuth ``row-cal'' scans of
all channels across RCW38.  The nominal feed separation is $18^\prime$.  These daily scans
are also performed with the telescope rotated by $60^\circ$ about the optical axis with
similar results.  \label{fig:beam_offset_rowcals}}
\end{figure}
\begin{figure*}
\begin{center}
\hspace{-0.25in}
\includegraphics[width=7in,clip]{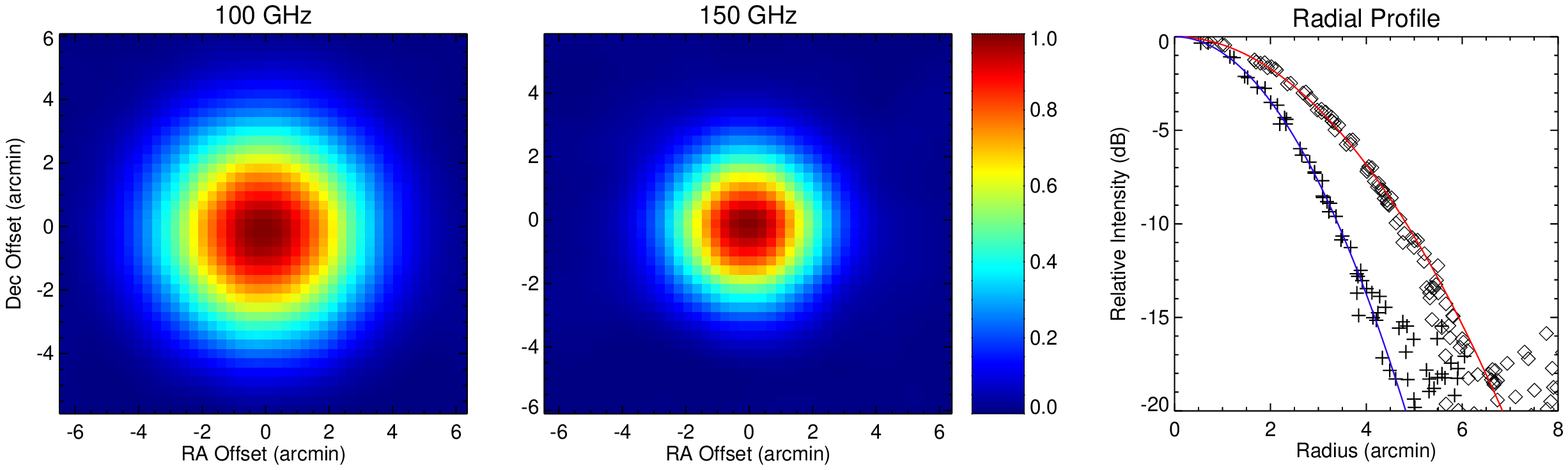}
\end{center}
\caption{The average QUaD beam at each frequency obtained by co-adding all detectors over three days of beam mapping runs on the source PKS0537-441. The beams have been smoothed with a $1.2^\prime$ FWHM Gaussian. (right) Radial profile of the un-smoothed data to the left and the best-fit Gaussians.\label{fig:quad_quasar_beams}}
\end{figure*}
QUaD used a nine parameter pointing model with the telescope control computer
handling full conversion from RA/Dec/paralactic angle request values to azimuth/elevation/deck
encoder command values.
The model parameters were established from a combination of optical data from a small
telescope attached to the elevation structure of the mount and from several special day-long
radio pointing runs.
These radio pointing runs consisted of scanning the central pixel across five bright, compact millimeter
sources (MAT6A, NGC3576, IRAS1022, RCW38, and IRAS08576) in both azimuth
and elevation, in a pattern known as a ``pointing cross.''
This determines the pointing offset between the optical and radio systems,
the flexure with elevation of the radio pointing, and the magnitude and angle
of the offset between the radio pointing direction and the rotation axis of the mount's third axis.
The azimuth axis tilts and encoder zero points were measured a few
times per season from the optical data.
Later radio pointing runs established the accuracy of the absolute pointing at $\sim 0.5^\prime$.

Approximately one day per month was devoted to making full raster maps using a bright source, usually RCW38 (Figure~\ref{fig:fp_plots}, top).
Gaussian profiles were fit to the raster data and were then used to determine the pointing offset of each feed.
The feed offsets obtained from twelve full beam maps are shown in Figure~\ref{fig:fp_plots} (bottom).
The offset angles used in the data analysis were determined by averaging over the twelve mapping runs,
to reduce the effects of the random pointing wander that occurs during each 16 hour run.
The RMS scatter in the feed offset angles versus the mean is $\sim 0.3^\prime$ (Figure~\ref{fig:rcw38_beam_centroid_histograms}).
There is no evidence for a systematic change in the feed offset angles over time.

In addition to the monthly raster scans in which the entire focal plane
was mapped, pointing checks were performed every eight hours during
routine observations with a pointing cross observation on RCW38
(Figure~\ref{fig:cross_obs_reduction.eps}).
The scatter of the derived pointing from the cross observations is $\sim 0.4^\prime$ (also in
Figure~\ref{fig:cross_obs_reduction.eps}) provides a useful cross-check
on the pointing uncertainty derived from the full focal plane raster
scans. Attempts were made to use these offsets to make pointing corrections
during offline data analysis but it was not possible to demonstrate any clear improvement using this
method.

Furthermore, in a procedure known as the ``row-cal,'' each of the 7 rows of the array is scanned in turn across RCW38 in azimuth.
Fitting one dimensional Gaussian profiles to the RCW38 blips in the row-cal time-ordered data of each channel allows us to monitor the relative angular separations between feeds in a given row.
Figure~\ref{fig:beam_offset_rowcals} shows the separations between pairs of adjacent feeds measured from one row-cal per day.
The scatter in these measurements is less than $0.1^\prime$ with little drift, confirming the long term stability of the array.

\subsection{Beam Characterization}\label{sec:beams}
Because of the geographic location, and the elevation limit imposed by
the ground shield, QUaD is not able to view planets.
As they are very bright and compact, they are the preferred targets for beam mapping
at millimeter wavelengths.
RCW38, which is the brightest of the compact HII regions visible to QUaD, has a
$\sim~1^\prime$ intrinsic width \citep[][and our own measurements]{coble2003} and rich extended
structure that makes it difficult to use for absolute determination of
beam parameters.
Because of this, the QSO PKS 0537-441 (WMAP source PMN
J0538-4405), was used for this purpose. It is a true point source,
but at only $\sim$~5\,Jy, compared to the 145\,Jy flux of RCW38 at 150 GHz,
the quasar observations
required significantly more integration time than was used for the
RCW38 raster maps. Figure~\ref{fig:quad_quasar_beams} shows a quasar
beam map obtained from three days of observation, averaged over all detectors
used for the final CMB maps.
The beam widths used in {\clempaper} ($5.0^\prime$ at 100 GHz and $3.5^\prime$ at 150)
are based entirely on maps of this quasar.
There is evidence for small channel-to-channel variations in width, and for ellipticity $< 10$\%.

\begin{figure}
\begin{center}
\includegraphics[width=3in,clip]{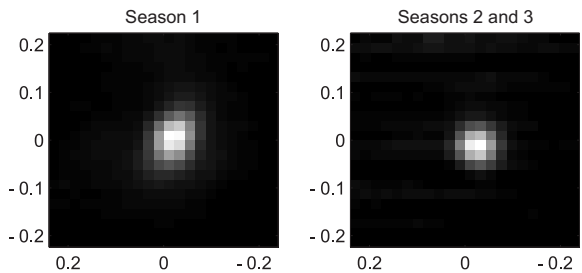}
\end{center}
\caption{Single channel beam maps from RCW38.
(left) A warp in the primary mirror made the beams elliptical during the first season.
(right) A shaped secondary mirror was installed for the second and third seasons, reducing
the average ellipticity to $\sim 5$\%. The axis units are degrees. \label{fig:beam_comparison}}
\end{figure}

\begin{figure}
\begin{center}
\includegraphics[width=3.4in]{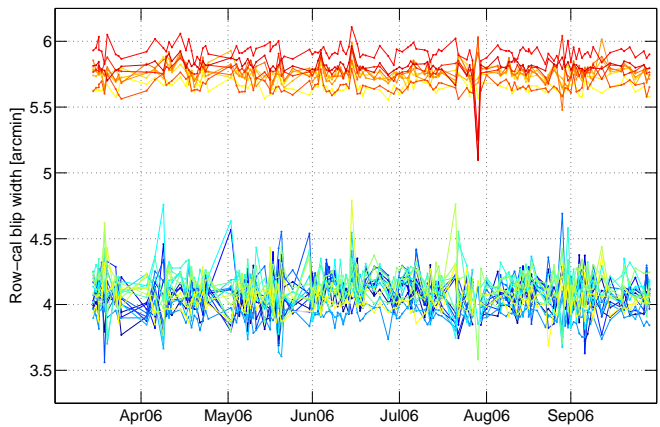}
\end{center}
\caption{The beam width of each feed horn, as measured from daily ``row-cal''
scans over RCW38.
Azimuth scans are made at seven elevation offsets from the source, corresponding to the center
of each row of feed horns in the focal plane.
Gaussian profiles are fit to the blip in the time stream corresponding to the source.
Because full raster maps are not made, pointing wander results in errors in the derived beam widths.
The larger scatter at 150\,GHz results from the higher
atmospheric noise present in this band.
There is no evidence for long
term drift in the beam width over the course of a season.
 \label{fig:beam_width_rowcals}}
\end{figure}

In the first season of observation, as shown in Figure~\ref{fig:beam_comparison},
the beams had significant ellipticity that was found to vary with external temperature as the
foam cone supporting the secondary expanded and contracted. The
ellipticity was traced to a small saddle-shaped warp of the
primary mirror. The shape of the primary was accurately measured after
installation on the QUaD mount, using an articulated measurement
arm.\footnote{Romer Series 3000i. www.romer.com}  The measured warp
has an angular dependence of $cos(2\theta)$ and an amplitude of
approximately 0.25\,mm at the outer edge of the mirror (where the beam
intensity is down by a factor of one hundred).

The beam shapes measured in season one can be accurately described by
an optics model that included the measured warp, and in the first
season this model was used to determine the secondary mirror position
that minimized the average ellipticity across the focal plane. In the
summer following the first season, a shaped secondary mirror was
installed that corrected for the primary, providing nearly symmetric
beams for the second and third season's observations. The optics
modeling is described in more detail in \citet{osullivan2008}.

Analysis of the row-cal scans of RCW38 from the first season showed that the beam
shape varied with the external air temperature.
A beam model, derived from the row-cal data, was used to correct for this effect.
For the second and third seasons, with the shaped secondary mirror installed, a similar analysis
showed no significant variation with temperature or time.
The beams have long-term stability well within the overall uncertainty
and no correction for time variation is applied (Figure~\ref{fig:beam_width_rowcals}).

Two polarization-specific systematics are present in the QUaD beams.
The first, known as squint, is an offset in the centroids of the beams from the two orthogonal detectors in a PSB pair.  The effect is most severe for the outer ring of 150\,GHz pixels, with a maximum offset of $0.2^\prime$.
It is stable over time and is dealt with during analysis by
using the measured offsets for each PSB during simulation runs.
The second effect is a mismatch in the beam shapes between the two beams of a PSB
pair.  Such effects can cause temperature anisotropies to appear as
false polarization signals. However, simulations show that at QUaD's
sensitivity, the measured level of beam mismatch is insignificant.
There are several possible causes for these systematics, including birefringence
or differential reflections at the lenses, multi-path through the AR coatings,
or polarization-dependent interactions with edges of optical elements.
Further study is underway to elucidate these effects.
For more detail on pointing and beam shape measurements see
\citet{Zemcov2006PhDT}.

\begin{figure}[t]
   \hspace{-0.3in}
\includegraphics[width=3.75in,clip]{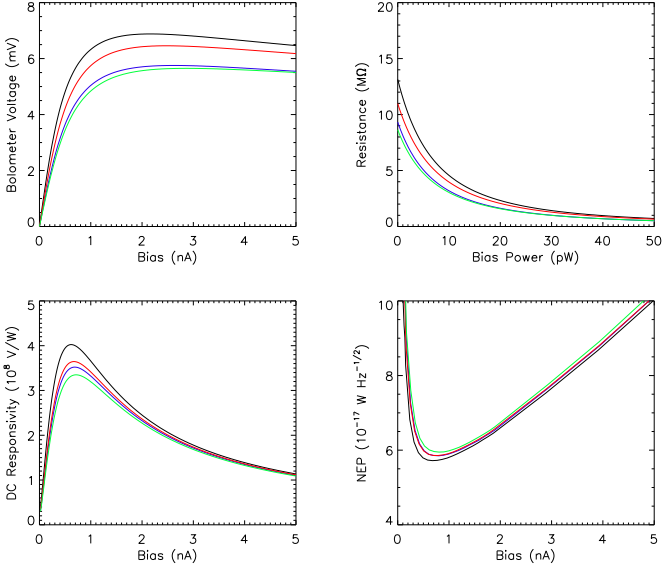}
   \caption{Load curves for four 150-GHz QUaD detectors taken on May 31, 2005 plotted in different units.  For low bias current, bolometers obey Ohm's law, but increased bias results in self-heating that decreases resistance.
These load curves were taken as part of a routine calibration set during CMB observations
at an elevation of 48~degrees and a baseplate temperature of 255\,mK.  During observations,
the detector bias current is set at 1.2 nA.} \vspace{0.2in}
   \label{fig:loadcurves}
\end{figure}

\subsection{Bolometer Parameters}
\label{sec:loadcurves}
Bolometers detect incident optical power by measuring the temperature increase of an absorber that is
in weak thermal contact with a fixed temperature bath.
In the QUaD detectors, the absorber temperature is measured with an NTD Germanium thermistor that has
a temperature-dependent resistance given by $R(T) = R_0\exp{\sqrt{\Delta / T}}$.
The thermal conductivity between the absorber and the bath is well described by
a power law $G(T) = G_0\left(T/T_0\right)^\beta$, where $T_0$ is an
arbitrary reference temperature, here taken to be 300\,mK.
Knowledge of these parameters is useful to convert measured bolometer voltages
into values of total absorbed optical power by solving the power balance equation
\newcommand{\Tbolo}{ T_\ns{bolo} }
\newcommand{\Tbath}{ T_\ns{bath} }
\begin{equation}
P + Q \; = \; \int_{\Tbolo}^{\Tbath}\;G(T)\;dT
\end{equation}
which states that in thermal equilibrium the sum of the electrical bias power, $P$, and the optical power, $Q$,
is equal to the power flowing from the absorber (at $\Tbolo$) to the bath across the weak thermal link, $G$.
It should be emphasized that this analysis is not needed to interpret the
astronomical data, because during normal observation the bolometers are biased so that
small changes in incident optical power result in proportional changes in signal voltage as in Equation~\ref{eq_v_psb}.

The bolometer parameters can be determined from a series of bolometer voltage
versus bias current measurements, known as load curves (Figure~\ref{fig:loadcurves}),
a procedure that is well documented in the literature \citep{sudiwala}.
Dark load curves, taken when the bolometers are blanked off, are particularly valuable for this process.
Some of the QUaD detectors were tested dark in a separate cryostat prior to installation.
All the detectors were dark tested with a special cool down in the QUaD cryostat during the Austral summer
between the first and second observing seasons.
The QUaD detectors were found to have typical values of $R_0, \Delta, G_0, \beta$ of $94\,\Omega$, 42\,K,
120\,pW/K, and 1.4 respectively.

As discussed in Section~\ref{sec:loading}, increased optical
loading warms the bolometers, causing the receiver gain to
fluctuate in response to changes in elevation or atmospheric conditions.
In order to increase the gain stability, Germanium bolometers are usually operated
at a bias that is somewhat higher than the most sensitive setting.
For this reason, the QUaD bolometer biases were set at
$\sim 1.2$\,nA, which is $\sim 40$\% larger than the bias setting that would give the lowest noise
equivalent power NEP (see Section~\ref{sec:sensitivity}).
This results in a roughly 30\% increase in gain stability
versus optical loading with only a $\sim 3$\% increase in NEP.
The typical DC responsivity at this bias setting is $\sim 3 \times 10^8$\,V/W.

\subsection{Spectral Bandpass and Optical Efficiency}
\label{sec:fts} \label{sec:optical_efficiency}
The spectral bandpass was measured for each channel using a Fourier
Transform Spectrometer.  Figure \ref{fig_avg_spectral_bands} shows the
average spectral bands for the two frequencies.  The band center is
computed as
\be \nu_0 = \frac{\int \, \nu \, f(\nu)\,d\nu}{ \int\,f(\nu)\,d\nu },
\ee
where $f(\nu)$ is the un-normalized transmission function as measured
by the FTS.
The bandwidth is computed as the separation of the half-power points in the transmission spectra.
The average measured center frequencies for the two bands are
$149.5\pm0.5$\,GHz and $94.6\pm0.6$, respectively, where the error indicates the
standard deviation of the values for all operating channels of the band.
The average bandwidths are
$41\pm2$\,GHz and $26\pm2$ for the 150 and 100\,GHz bands, respectively.

A bolometric receiver that is background limited has a sensitivity that
is inversely proportional to the square root of the optical efficiency.
The optical efficiency of the QUaD receiver was determined from load
curves taken during laboratory observations of beam-filling blackbody
loads made from unpainted Eccosorb CV-3 absorbing foam. The measurement
entails recording load curves with the instrument looking at a room temperature
load and then at a liquid nitrogen load. The additional electrical
bias power needed to warm a detector that is observing the cold
load to the temperature that it was at when observing the ambient load
gives the difference in optical power actually received from these two
sources. A comparison with the expected difference in optical power for
a perfect system with an optical efficiency of 100\% gives the optical
efficiency of each detector.  The mean receiver efficiency was measured to be
$34\pm6$\% at 150\,GHz and $27\pm4$\% at 100\,GHz where the error again indicates the
in-band channel-to-channel scatter.

\subsection{Detector Time Constants and Transfer Function} \label{sec:tau}
\begin{figure*}
   \begin{center}
\includegraphics[width=7in,clip]{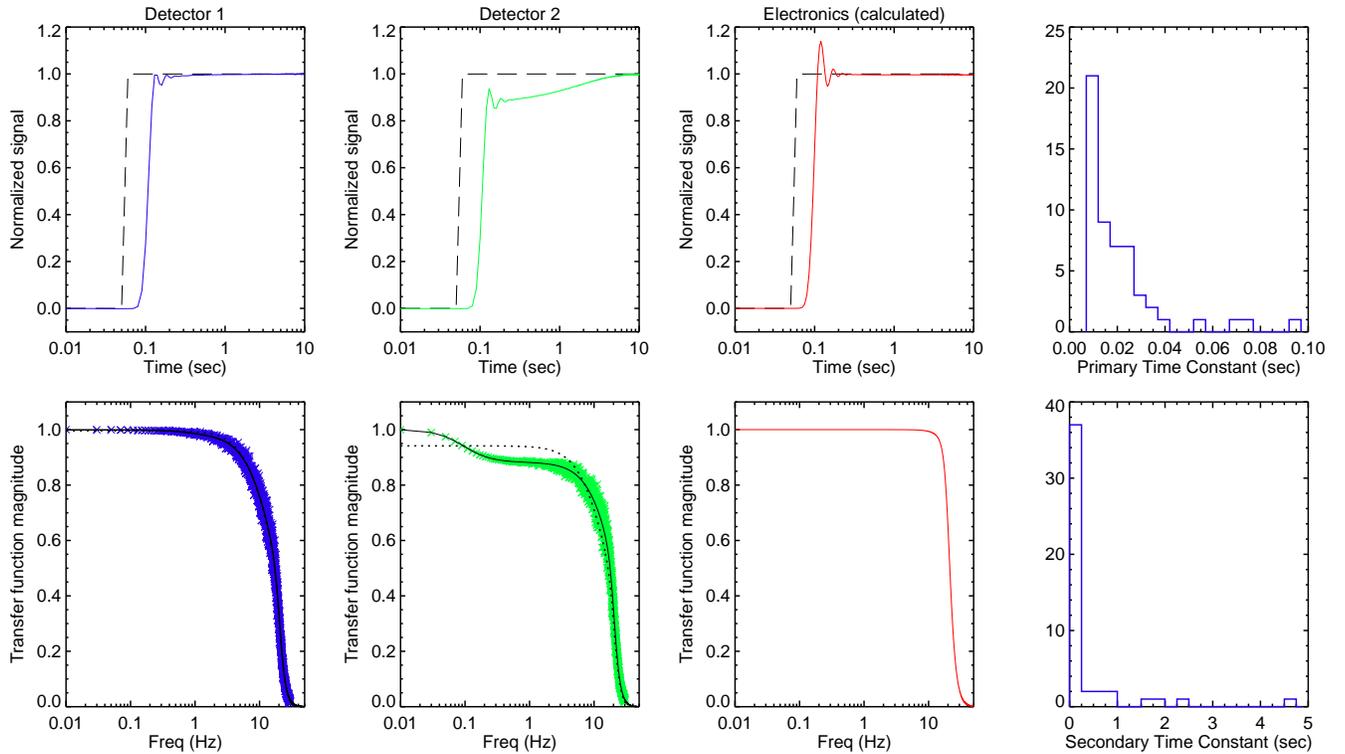}
   \end{center}
\caption{A Gunn diode source was used to supply a square wave signal with fast edge transitions to the focal plane in order to measure the detector time constants.  (left) The Gunn diode reference signal (black dashed) and the measured step response of two detectors (blue and green).  The initial $\sim 150$\,ms is dominated by the electronics transfer function which introduces the overshoot and ringing (red).  The transfer function for the two detectors, derived from Fourier transforming the time domain data, is shown below the time ordered data.  Fits to a single time constant model (dotted) and a dual time constant model (solid) are shown. The two right hand panels show the distribution of both the fast (upper right) and slow (lower right) time constants for all of the QUaD detectors.}
   \label{fig:gunn_tc_data}
\vspace{0.1in}
\end{figure*}
The system transfer function must be well understood in order to
convert measured detector voltage time streams into maps of intensity on
the sky. For QUaD, the overall transfer function is the product of the
bolometer response, $H_b(\omega)$ and the readout electronics response,
$H_e(\omega)$.  The electronics transfer function can be
calculated from the circuit design and is given by
\begin{eqnarray} \label{eq_def_H_e}
 H_e(\omega) &=& H_1(\omega) \cdot H_2(\omega) \\
 H_1(\omega) &=& \frac{1}{(s_1^2 + A s_1 + 1)(s_1^2 + \sqrt{2} s_1 + 1)(s_1^2 + s_1/A + 1)} \nonumber \\
 H_2(\omega) &=& \frac{1}{(s_2^2 + \sqrt{2}s_2 + 1)} \nonumber
\end{eqnarray}
where $A=-2 \cos(7\pi/12)$, $s = i\omega / (2\pi f_{3\mbox{\scriptsize dB}})$
and the 3\,dB points are $f_{3\mbox{\scriptsize dB}}=20$\,Hz  for $H_1$
and 30\,Hz for $H_2$. This transfer function was verified by laboratory
measurements and deviations from the model were found to be negligible
over the frequency range of interest for CMB measurements (0.1-2\,Hz).

The bolometer response dominates the electronics response in the
science band and accurate measurements of $H_b$ are essential in order
to correctly measure the CMB power spectrum. The response function of an
ideal bolometer is $S(\omega) = S_{\mbox{DC}} / (1 + i \omega \tau)$
where $S$ is the detector responsivity (Volts / Watt), and the time
constant $\tau = C/G$ where $C$ is the heat capacity of the absorber
and $G$ is the thermal conductivity between the sensor and the bath.
Values for $\tau$ of order 30\,ms are typical of QUaD detectors although
there is substantial device-to-device variation. Both $C$ and $G$ are
temperature dependent, so it is essential to measure time constants
with similar optical loading and electrical bias power as during
observation.

Two methods were used to determine the time constants for the QUaD
detectors. The first method involved scanning the telescope
back and forth in azimuth over RCW38 and recording the shift
in the apparent position of the source for the
forward and backward scan directions. The second method involved
a special test run during the Austral summer.
The telescope was illuminated with a Gunn diode RF source that was
chopped with a slow (100 second) square wave.
This measures the system step response from which the transfer function can be
derived (Figure~\ref{fig:gunn_tc_data}). Care was taken to ensure that the Gunn setup did not
substantially increase the optical loading on the detectors.
The results of the two methods are in excellent agreement
with each other; however, the Gunn measurements offered a much
larger signal-to-noise ratio, and these are the values used in \clempaper.

It was found that for about half the detectors the
single time constant model is a poor fit to the true response. The
addition of a second time constant term, so that the overall detector
transfer function is modeled as
\be
H_b(\omega) = \frac{1 - \alpha}{1 + i \omega \tau_1} + \frac{\alpha}{1 + i \omega \tau_2} \label{eq_def_H_b}
\ee
where $\alpha \le 0.5$, was found to accurately model the detector
response in all but two cases --- these detectors were not used.
In some cases the second time constant is found to be extremely long ($\tau_2 > 1$~second), but is
still very well fit by the double time constant model.
These detectors are used in the science analysis.
The  physical origin of the second time constant is believed to be debris on the PSB
which is in weak thermal contact with the absorbing grid. Figure~\ref{fig:gunn_tc_data} shows the transfer functions for two detectors,
and the distribution of time constants for all the detectors used in the season 2/3 analysis.

\subsection{Polarization Angles \& Cross-polar leakage}\label{sec:pol}
\begin{figure*}
    \begin{center}
\includegraphics[width=7.0in]{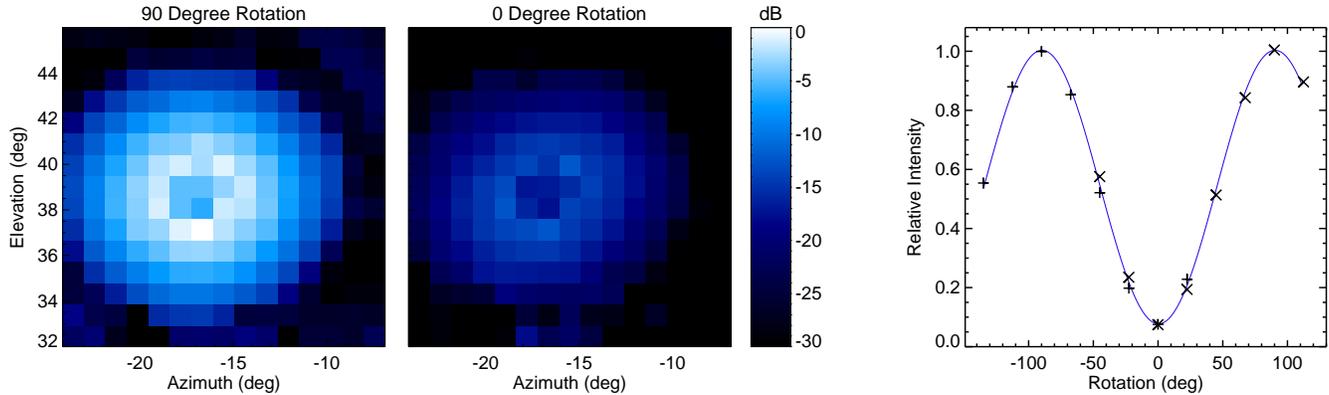}
    \end{center}
    \caption{(left) Maps of the near field response of a single QUaD pixel made by rastering the telescope over a polarized calibration source. The left and middle panels show the response with the relative angle of the telescope deck and the source aligned for maximum and minimum response, respectively.
    (right) The signal from each raster map is integrated to provide a single data point on this plot.  The best-fit sine wave gives the cross-polar leakage and PSB orientation angle.  The `+' data points were taken with the source grid vertical, and the `x' data points with the grid horizontal.  The rotation of the telescope was limited to $\pm 80$ degrees to prevent cryogens from spilling.
      \label{fig:polcal_rasters}}
\end{figure*}
\begin{figure}
    \centering
\includegraphics[width=3.4in]{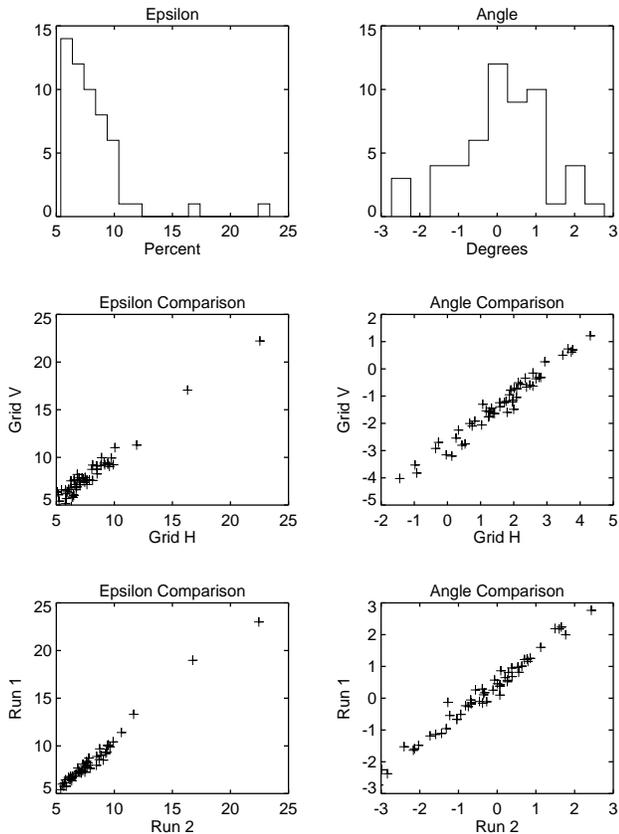}
    \caption{Upper panels: the distribution of the cross-polar leakage $\epsilon$, and the deviation of the PSB angle $\theta$ from the nominal angle, for each pixel.
    The two middle panels compare the derived $\epsilon$ and $\theta$ values for the source polarization direction oriented vertically and horizontally, to show that source effects are small.
    The two lower panels compare the $\epsilon$ and $\theta$ values derived from two different calibration runs two months apart, showing that the polarization properties of the instrument are stable with time.
      \label{fig:polcal_stats}}
\end{figure}
The voltage response of a single PSB to arbitrarily-polarized
incident radiation is given by
\be
v = s\bigg[\istk \: + \: \frac{1-\epsilon}{1+\epsilon}\left(\qstk \cos 2\theta + \ustk \sin 2\theta
                \right)\bigg] \label{eq_v_psb}
\ee
where $\theta$ gives the orientation angle of the PSB, $\epsilon$ gives
the cross-polar leakage and $s$ is a calibration constant that depends
on detector responsivity, optical throughput ($A\Omega$), optical
efficiency, bandwidth, and readout electronics gain
\citep{2003SPIE.4855..227J}. From equation \ref{eq_v_psb}, leakage can
be seen to result in a loss of optical efficiency to polarized
radiation. This has the effect of reducing sensitivity (Section~\ref{sec:sensitivity})
but not of mixing Stokes parameters. The
measurement of $\epsilon$ and $\theta$ is described in this section.
The determination of the calibration constant, $s$, is described in
Section \ref{sec:calibration}.

Because there are no bright, well calibrated polarized astronomical
sources visible at these wavelengths, we characterized the polarized
response of the instrument by observing a near-field, polarized
source mounted on a tower located outside the ground shield. The source
consisted of a rotating chopper wheel viewed through a circular
aperture and a linear polarizing grid.\footnote{The polarizing grid,
manufactured in Cardiff, is an array of copper traces
at a 10 micron pitch supported on a thin polypropylene sheet. At
150\,GHz, it was measured to have polarization efficiency of greater
than 99\%.} The chopper wheel alternated between ambient temperature
Eccosorb and reflective aluminum aimed at cold sky with a
modulation rate of 5\,Hz. A phase-synchronous reference signal was
digitized on one of the spare analog input channels and used to
demodulate the bolometer voltages in post processing.

A series of $15^\circ \times 15^\circ$ raster maps of the calibration source were recorded,
rotating the receiver about the optical axis in steps of $22.5^\circ$ between each map.
After eight maps, the source grid was rotated by $90^\circ$ and the entire process was repeated.
Each raster map required approximately 45 minutes of observation.
Two complete polarization calibration runs were performed two months apart, during the second season of observation.
For the first run, a source aperture made from aluminum was used. For
the second run, an Eccosorb aperture was used.   Results from the two
runs are similar, with slightly lower values of $\epsilon$ obtained on the second
run due to less depolarizing effects from the source aperture.

The signal from each raster map is integrated to provide a single data
point on a plot of measured signal versus relative grid / PSB
orientation angle (Figure~\ref{fig:polcal_rasters}). These data points
are then fitted to a sine wave. The cross-polar leakage, $\epsilon$, is
given by the difference between the sine wave minimum and the $x$ axis.
The absolute orientation angle of the PSB is given by the phase angle
of the sine wave.

Figure \ref{fig:polcal_stats} shows the results of these two
polarization calibration runs. We find that the average cross-polar
leakage of all the channels that are used in the season 2/3 analysis is
0.08. In order to check for systematics in the measurement of
$\epsilon$ and $\theta$, the results from the two calibration runs can
be compared, and additionally each calibration run can be split into
two halves, since half of the data points were taken with the source
grid aligned horizontally and half vertically. By comparing the results
obtained from independently analyzing the ``grid horizontal'' and ``grid
vertical'' data subsets, and the results from the two separate calibration
runs, an uncertainty on the mean value of $\epsilon$ of $\sim \pm 0.02$ is
obtained.

The measurements of cross-polar leakage are consistent with laboratory
measurements of individual pixels made using a small testbed cryostat
prior to integration of the QUaD receiver. The optical system of the
test cryostat is simpler than in the full receiver, containing only
filters and a vacuum window and no lenses. Multiple feeds and detectors
at each frequency were characterized in the testbed. This suggests that the lenses and
the telescope are not significant sources of cross-polar leakage.

The RMS scatter of the PSB angles about their nominal orientation was
found to be $1.2^\circ$, and the deviation from orthogonality within a
pair was similarly of order $1^\circ$. Several mechanisms contribute to the errors in
PSB orientation including machining tolerances of the focal plane plate
and PSB modules, transmission through the cryostat optical chain, and
the alignment of the absorber within the PSB module.
Simulations show that random offsets in the PSB angles at this level average down
and have no effect on the resulting CMB polarization power spectra.

A larger ($\sim 2^\circ$) systematic uncertainty in the absolute position
of the PSB orientation angles results from differences obtained in the two
calibration runs and from separate fits to the source grid horizontal and source grid vertical data subsets.
Systematic rotation of the detectors is more problematic than random scatter
since it does not average down when detector signals are combined.
Misalignment produces leakage of the dominant CMB E-mode power into the much
weaker B mode signal.
However, simulations show that at QUaD's sensitivity a rotation of order $10^\circ$ ---
five times larger than the actual uncertainty --- is necessary to
produce a detectable amount of spurious BB signal.

\subsection{Optical Loading}\label{sec:loading}
\begin{figure*}[t]
   \begin{center}
\includegraphics[width=6.5in,clip]{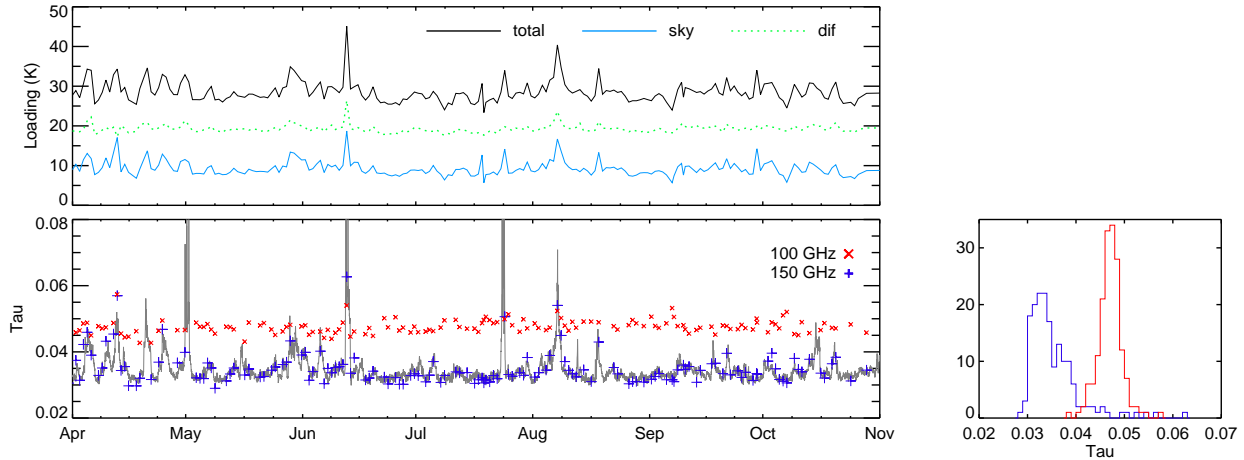}
   \end{center}
   \caption{
(top) Average Raleigh Jeans optical loading temperature for the 150 GHz
channels, measured during the 2006 observing season. The total loading
(black), is derived from daily load curves and the atmospheric loading
(light blue) is derived from daily skydip measurements. The difference
(dotted green) remains flat over the season, indicating that the
loading is not increasing due to, for example, snow accumulation on the
foam cone. (bottom) Atmospheric optical depth (tau), as measured by
daily QUaD skydips for both observing bands. Data from the AST/RO $350\, \mu$m tipper (located in a neighboring building) is shown in gray. The
gray line shows an extrapolation of the $350\, \mu$m optical depth from
the tipper data to 150\,GHz using the relation in the text. There is
excellent agreement between the two.  (right) Histogram of the 100 (red) and
150\,GHz (blue) tau measurements.  \label{fig:atmospheric_tau}}
\vspace{0.2in}
\end{figure*}

Optical loading on the detectors arises from thermal emission in the
instrument, from atmospheric emission and from astronomical sources,
including the CMB. The total loading is:
\newcommand{\tfixed}{T_\ns{telescope}}
\begin{equation}
 T_\ns{load}(\theta) = T_\ns{receiver} + T_\ns{telescope} + (1 - \trans)\; T_\ns{atm} + \trans T_\ns{CMB}
 \label{eq_sd_tload}
\end{equation}
where $T_\ns{atm}$ and $T_\ns{CMB}$ are the Raleigh-Jeans temperatures
of the atmosphere and the CMB. The loading per detector averages $\sim 30$\,K
(Table~\ref{tab:noise_budget}).

Excess loading reduces sensitivity by increasing the photon noise contribution and by warming the detectors, which reduces
responsivity (Section \ref{sec:sensitivity}).
The QUaD receiver was designed to minimize internal loading.
All transmissive components including the window, lenses, and filters are anti-reflection coated with thin layers
of intermediate index material. Also a blackened cold stop located on
the 4\,K stage intercepts the sidelobes from the feed horns.

For QUaD, the atmospheric transmission is measured daily with a
procedure known as a skydip, during which the detector voltages are recorded as the telescope is tipped in elevation
from zenith to $45^\circ$.
The atmospheric transmission depends on zenith angle as $\trans = e^{-\tau / \cos \theta}$ where the
optical depth, $\tau$, depends on the frequency band of observation.
The combination of the excellent atmospheric conditions at the South
Pole, with $\tau$ at 150\,GHz less than $\sim 0.04$, and the limited elevation
range accessible to QUaD means that
$\trans \approx 1 - \tau / \cos \theta$ and the total loading temperature is given by
 \be
  T_\ns{load}(\theta) \approx T_0 + \frac{\tau \, T_\ns{atm}}{\cos\theta}.
 \label{eq_tau_approx}
 \ee
Consequently it is not possible to separately fit for $\tau$ and
$T_\ns{atm}$ from QUaD skydip data alone. The degeneracy was broken by
using data from the AST/RO $350\, \mu$m tipper which is located on an
adjacent building and which measures $T_\ns{atm}$ (and $\tau$ at
$350\, \mu$m) several times per hour. By using the tipper measurement of
$T_\ns{atm}$, the value of $\tau$ at the QUaD observing frequencies can
be derived.\footnote{The atmospheric temperature, $T_\ns{atm}$, is a
weighted, line-of-sight averaged temperature and is not, in general,
the same as the surface air temperature.} During routine observing,
QUaD performs one skydip per day.

Figure \ref{fig:atmospheric_tau} shows the measured $\tau$ values for
the 2006 season. It can be seen that the derived optical depth is
larger at 100\,GHz than at 150\,GHz, as expected from the atmospheric
transmission model in Figure \ref{fig_avg_spectral_bands}. The optical
depth is, however, more variable at 150\,GHz than at 100\,GHz because
the former is more sensitive to the water vapor content. A comparison
of the $350\, \mu$m tipper data and the 150\,GHz optical depth derived
from QUaD skydips gives a best-fit linear relationship of
\be
\tau_{150} = 0.020 + 0.0127 \times \tau_{350\mu\mbox{\scriptsize m}}
\ee
and shows that the two are strongly correlated. 

\section{Rejection of Spurious Signals}  \label{sec:interference}
All bolometric detector systems are susceptible to contamination from a
variety of effects including stray light, crosstalk, radio frequency
interference (RFI), and microphonic pickup. Although the differential
nature of polarization measurements provides an extra level of
rejection compared to total-power measurements, careful design is still
required. To monitor any residual effect, the focal plane includes
several dark bolometers and fixed-resistors.

\subsection{Stray Light}
The QUaD bolometer modules are not light tight, so a 250\,mK shield is
mounted beneath the focal plane bowl to prevent stray light from
entering through the modules themselves. All joints in the shield are
stepped to provide a convoluted path for light, and the inside is
blackened with carbon-loaded Stycast 2850FT. The volume enclosed by the
focal plane bowl and the shield is vented through small-diameter, teflon
tubes embedded in the absorber. The long tubes are bent and glued to
the inside of the shield to remove any light path while still providing
an exit for gas. Load curves of the \emph{dark} bolometers show
negligible stray light inside the focal plane enclosure. Correlations
of the output signals from the dark bolometers with the signal from
light channels made while the receiver observes a modulated optical
load show that cross-talk (both optical and electrical) is less than
one percent.

\subsection{Electrical Pickup and Radio Frequency Interference}
QUaD is well-protected from pickup in the warm readout electronics and
cabling. The electronics boxes mount onto RF-tight interface boxes
which themselves attach directly to the cryostat bottom flange and are
sealed with RF braid. All signals entering or exiting the interface
boxes pass through filtered D-Sub connectors (Section~\ref{sec:electronics}).
The signals are again filtered at the 4\,K bulkhead, this time with
filtered micro D connectors. The signal paths
for the two halves of each PSB module are kept as similar as possible,
so that any electrical pickup will affect each in a similar manner and
will be attenuated by differencing.

RF Interference in the low GHz range is particularly insidious as it
easily enters the cryostat through the window, where it is picked up by
the signal wiring and then coupled to the detectors.  Any RF power
picked up in this way will heat the thermistors, causing a decrease in
voltage that can be confused with true signal. Fortunately, the radio
frequency environment at the remote South Pole site is both
exceptionally clean and tightly controlled, especially during the
Austral winter. All communication and weather monitoring transmissions
are in the low MHz region of the spectrum and are thus blocked by the
size of QUaD's optical snout. During three years of observations, no
contaminating transmissions were detected by the instrument during the
winter observing season.

\subsection{Microphonic Pickup}
Mechanical vibrations in the high-impedance signal wiring between the
detectors and the JFET buffers can cause spurious signals -- an effect
known as microphonic pickup. To mitigate this effect, the JFET buffer
amplifiers are placed as close as possible to the detectors, reducing
the length of the susceptible wiring, and all the high-impedance wiring
is carefully tied down along its entire path to limit vibration.
Additionally, the QUaD mount has extremely smooth tracking;
accelerometer measurements showed the vibration level to be under
0.5x10$^{-4}$\,g Hz$^{-\frac{1}{2}}$ within the range of 40---200\,Hz
during typical azimuth scans. Finally, the phase compensation in the
lockin amplifiers is frequently adjusted which minimizes sensitivity to
microphonically induced phase shifts.

Although these measures significantly reduce the amount of microphonic
pickup, there remain some broad resonant features in the output
signals. However, the adjustable bias frequency allows operation in a
clean region so that the signal frequencies remain uncontaminated. With
the bias frequency properly adjusted, the receiver shows no increase in
white noise level or the $1/f$-knee frequencies, even during telescope
motion.

\begin{figure}
   \begin{center}
\includegraphics[width=3.25in,clip]{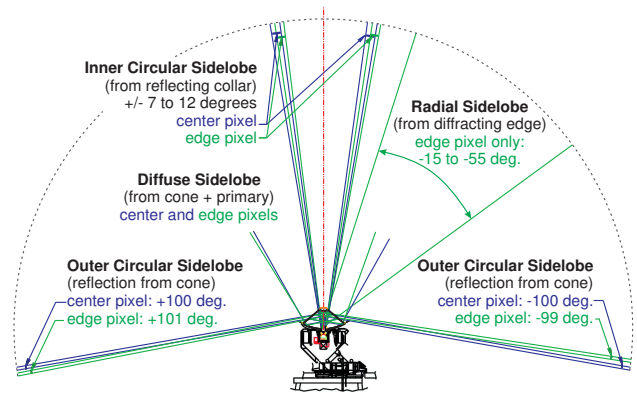}
   \end{center}
\caption{Schematic showing the location and origin of the telescope sidelobes.\label{fig:sidelobe_schematic}}
\end{figure}

\begin{figure}

   \begin{center}
\includegraphics[width=3.25in,clip]{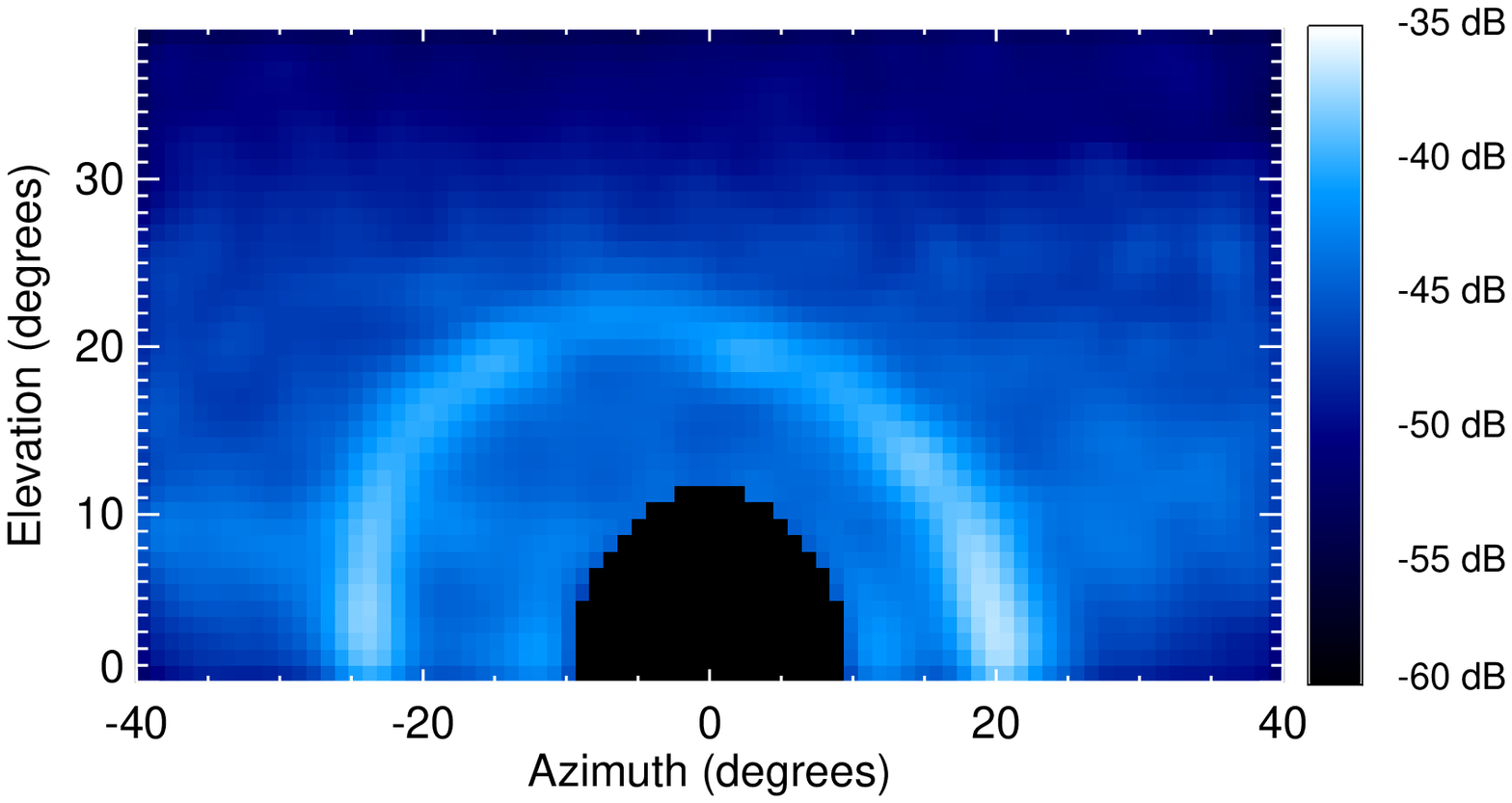}
   \vspace{0.075in}
\includegraphics[width=3.25in,clip]{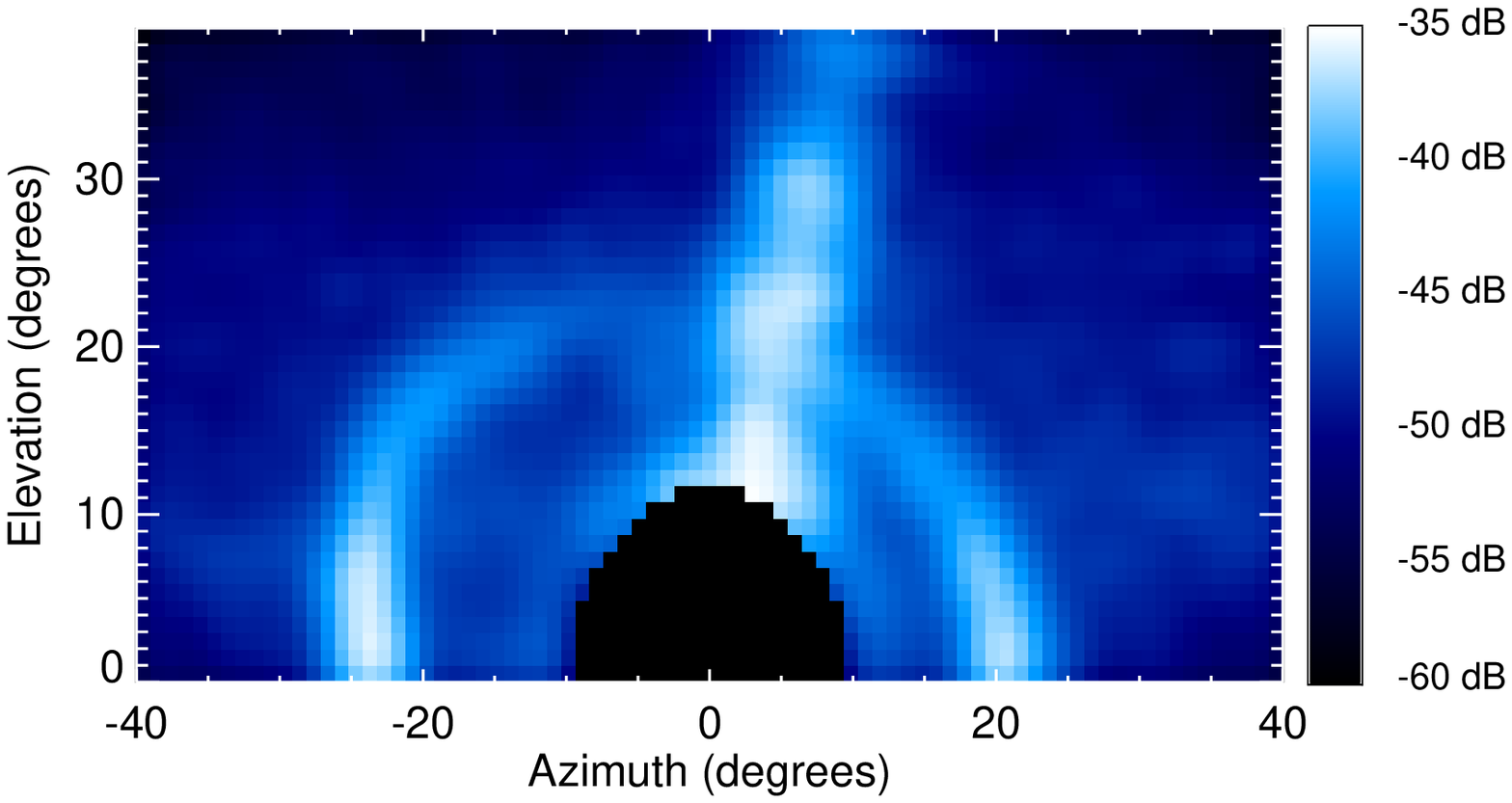}
   \end{center}
\caption{Sidelobe measurements on two 150\,GHz channels using a chopped Gunn source mounted on the
lip of the ground shield, at an elevation of $21^\circ$. Axis coordinates are offsets relative to the source.  The units are
dB per 0.5 square degree pixel, relative to the main beam.  The channel shown in the top plot is
from the inner group.  The bottom channel is from the outer group, and shows the characteristic radial sidelobe.
This data was taken after the first season, with the original parabolic collar baffle.}
   \label{fig:MZ_sidelobes}
\end{figure}

\subsection{Cosmic Rays}
\begin{figure}
   \begin{center}
\includegraphics[width=3.0in,clip]{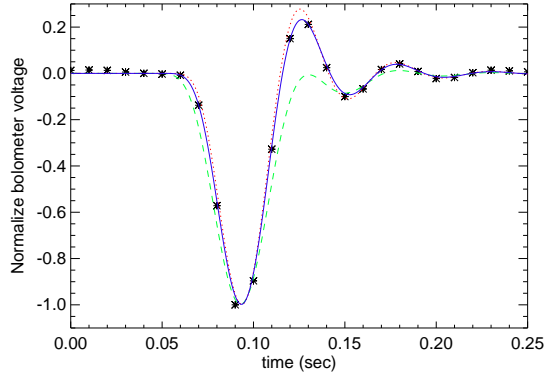}
   \end{center}
\caption{The response of a channel to a cosmic ray hit.  The crosses show the measured data points.  A good fit is given by the transfer function described in Section~\ref{sec:tau} with the detector time constant left as a free parameter (solid blue curve).  In this case, the best-fit detector time constant is 6\,ms, much shorter than the optically measured value of 19\,ms (dashed green).  The electronics-only impulse response is given by the dotted red curve.
\label{fig:cosmic_ray_impulse}}
\end{figure}
Cosmic ray hits cause a short glitch in the data stream of the affected channel \citep{woodcraft2003}.
See Figure~\ref{fig:cosmic_ray_impulse} for a typical cosmic ray event.
The PSBs used in QUaD have a small cross-section to cosmic rays, so these events occur infrequently.
During observation at the South Pole, QUaD experienced approximately 0.5 cosmic ray events per channel per hour.
These events are tagged in the low-level data reduction, and contaminated regions are excluded from further analysis.
This results in a loss of data of much less than one percent.

Cosmic ray impacts closely approximate delta function power inputs, providing a measure of the system impulse response function.
These events are typically well-fit using the transfer function described in Section~\ref{sec:tau}, with the detector time constant left as a free parameter.
The best-fit detector time constants are shorter than those obtained from optical measurements (on average, by a factor of three).
They vary for different events on the same device and show little correlation with the optically measured time constants.
These effects are due to poor thermal conductance across the PSB absorber combined with the random impact location of each event.
Cosmic ray glitches are thus not useful for measuring the detector time constants, but they do provide a monitor of the electronics impulse response.

\begin{figure}
   \begin{center}
\includegraphics[width=3in,clip]{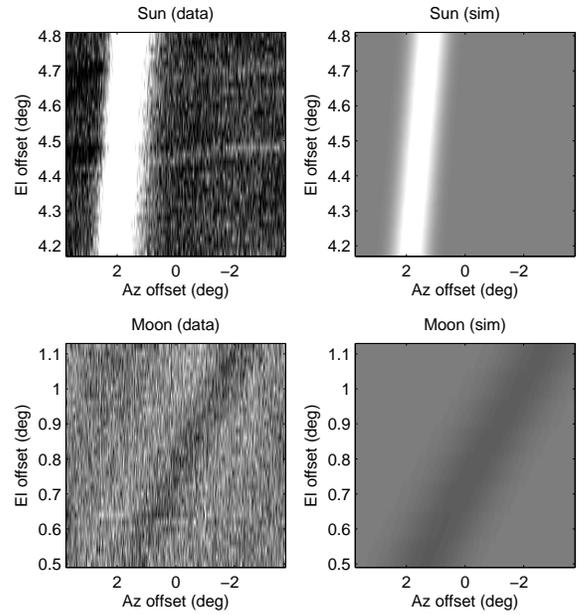}
   \end{center}
\caption{
(left) Single PSB pair maps of data contaminated by the sun and moon.
Note these sources move across the sky at different rates, resulting in
different stripe angles in the maps.
(right) Simulated maps that incorporate a
best-fit cone angle, the sidelobe width, and polarization
information based on the sun / moon coordinates during the same
day as the real data.  The cone opening angle (allowed to vary from
$96^\circ$ to $101^\circ$) is the only free parameter of this model.}
\label{fig:sunmoon_datasim}
\end{figure}

\begin{figure}
   \begin{center}
\includegraphics[width=2.75in,clip]{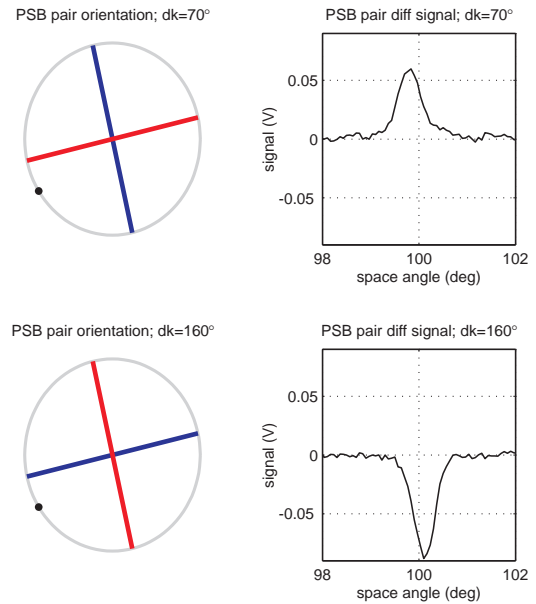}
   \end{center}
\caption{PSB pair orientation with respect
to the sun position in the frame of the focal plane, for two
telescope deck rotation angles separated by $90^\circ$. As expected, the
pair differenced (A-B) signal is positive when detector A (blue)
is perpendicular to the radial direction to the source. Likewise,
with the sun in a very similar position, the pair rotated by
$90^\circ$ produces a negative pair difference signal.}
\label{fig:pairdiff_suntod}
\end{figure}

\subsection{Sidelobe pickup} \label{sec:sidelobes}
Figure~\ref{fig:sidelobe_schematic} shows the location of the telescope sidelobes.
The inner circular sidelobe arises from reflection off
the cryostat collar baffle (Section~\ref{sec:optics}).
The radial sidelobe, caused by truncation of the beam inside the cryostat,
is present in the outer ring of 150\,GHz channels.
Finally, all channels exhibit an outer circular sidelobe at approximately $100^\circ$ from the
optical axis caused by reflection off the foam cone.
These sidelobes have been extensively studied with a series of special mapping runs, as
described in this section.

An optics model was used to estimate the amplitude of these sidelobes.
Of the power leaving a feed horn, approximately 5\% passes through the
hole in the secondary mirror.  Another 10\% is reflected from the cryostat
baffle to the inner circular sidelobe.
Combined, these two terms represent the loss of efficiency that results
from blockage by the on-axis secondary mirror.
A further $\sim 2$\% of the power is lost due to reflection off the foam
cone.
Half of this power forms the narrow, $100^\circ$ sidelobe, while the
rest scatters off the primary resulting in a wide, diffuse sidelobe
at $\sim 30^\circ$ off the boresight.

The inner circular sidelobe and the radial lobe were studied
with a series of special mapping runs using a chopped Gunn
diode mounted on the top of the ground shield as a source.
The results of these runs are shown in Figure~\ref{fig:MZ_sidelobes}.
After these tests, it was decided to replace the original first season
parabolic baffle with a flat, reflecting baffle.
This moved the sidelobe from $\sim 25^\circ$ off the optical axis
to $\sim 10^\circ$ for seasons 2 and 3, where it would be less likely to intercept a
contaminating source.
The total power in this sidelobe remains the same.

As a further test during the summer of 2006,
the annular cryostat baffle was coated in absorbing material,
and a cylindrical absorbing baffle was placed above the cryostat snout.
Tests with the Gunn source indicated that these baffles eliminated the inner circular
sidelobe and the radial lobe.
However, the increase in the average detector loading from these baffles resulted in
an unacceptable loss of sensitivity and they were removed
before beginning CMB observations.
With just the absorbing collar baffle, an average of 10\,K increased loading
was seen at 150\,GHz versus a 20\,K increase with both baffles.
There were channel-to-channel variations of order a factor of two.
This indicates that the inner circular sidelobe and radial sidelobe are
of similar strength.
These loading increases are slightly less than expected from the optics model.

The sidelobes can introduce spurious signals into the data via pickup from the ground
as the telescope scans or if they intercept a bright astronomical source.
Despite the reflective ground shield surrounding the telescope,
low-level ground pickup is seen in the raw data.
The ground signal appears at low frequencies and is very effectively
suppressed by high-pass filtering each scan.
Furthermore, a field differencing observation strategy was employed for all CMB observations.
Two neighboring CMB fields, separated by exactly 30 minutes in right ascension, were observed.
The ``lead'' field was observed first, and ``trail'' field was observed precisely
30 sidereal minutes later, with the telescope repeating an identical scan path with respect to the ground.
The difference field (lead - trail) is free of any
ground pickup that is stable on time scales of 30 minutes or longer.

Because of the low level of the sidelobes, only bright astronomical sources can create contamination.
The sun and planets are always below the ground shield during CMB observations.
Pickup from the galactic plane was simulated and found to contribute at a level
well below the sensitivity of QUaD.
This leaves the moon as the most problematic source.
The geometry of the site and field location confines the moon to a ``space angle'' range
of $\sim 75^\circ$ to $115^\circ$ measured with respect to the telescope boresight.
Therefore, moon contamination only enters the data via the $100^\circ$ sidelobe.

The moon, if it is appropriately positioned, leaves a characteristic stripe
in the maps, at an angle consistent with its rate of movement across the sky
(Figure \ref{fig:sunmoon_datasim}, left).
A model was developed that accurately predicts when moon contamination
will occur based on the relative position of the moon to the telescope (Figure \ref{fig:sunmoon_datasim}, right).
This model was used to cut any days with possible contamination, even if no
visible evidence was seen in the maps.

Although the foam cone was designed to have a nominal opening
angle of $100^\circ$, in practice this can vary due to thermal
expansion/contraction of the cone, inhomogeneities in the adhesive,
small tilts of the cone with respect to the boresight, and
to a lesser extent mechanical loading. These effects cause
the sidelobe space angle to fluctuate slightly as a function of
time and of telescope deck rotation angle.
We found that pickup from the moon in
the CMB data occurs over the range 96.5 $<$ space angle $<$ 101 degrees,
and this forms the basis for our very conservative
moon cuts as discussed in \clempaper.

\begin{figure}[t]
   \begin{center}
   \parbox{3.5in}{\hspace{-0.2in}
\includegraphics[width=3.7in,clip]{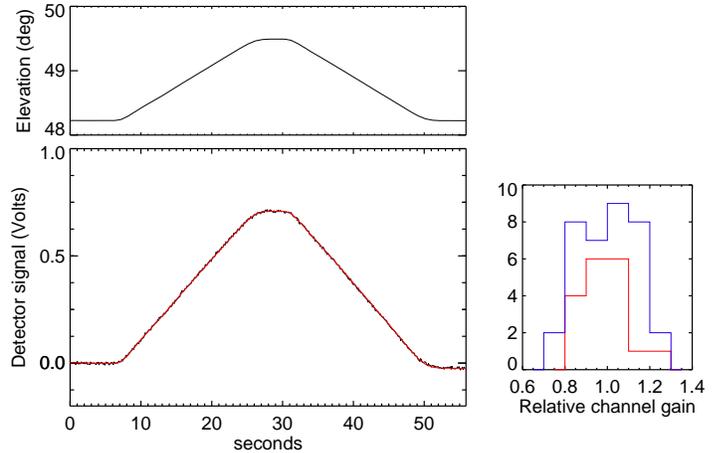}
   }
   \end{center}
\caption{Nodding the telescope in elevation (top) introduces a large signal in the amplified bolometer time stream (bottom) that is used to normalize the relative channel gains.  The best-fit model ($1 / \cos(\mbox{ZA})$ plus a linear drift) is shown in red.  The relative channel gains as derived from this el-nod for the 150 GHz (blue) and 100 GHz (red) channels are shown to the right.  This distribution results
from differences in optical efficiency and detector responsivity.
   \label{fig:elnod}}
\end{figure}

\begin{figure*}
\begin{center}
\includegraphics[width=2.5in,clip]{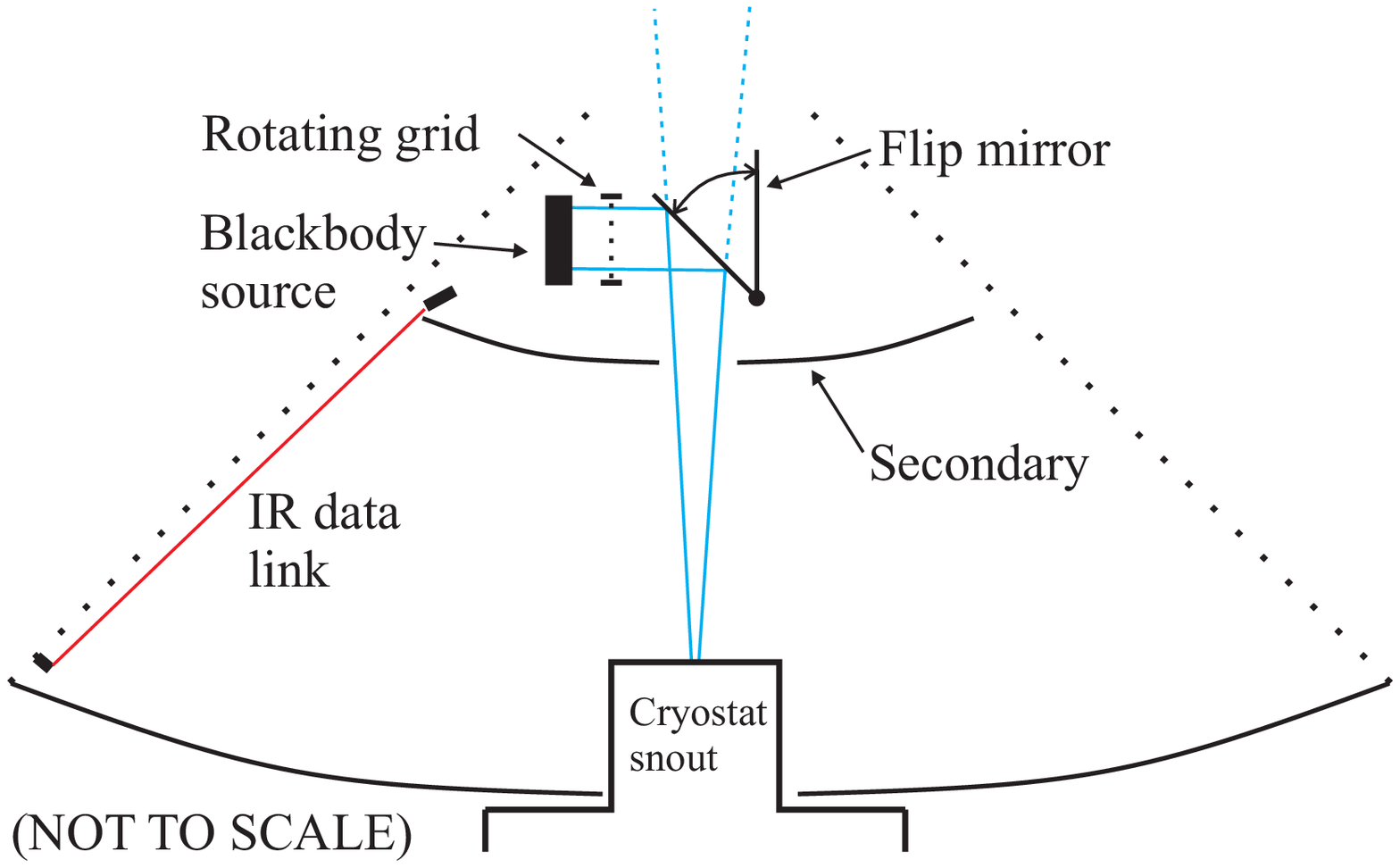}
\hspace{0.25in}
\includegraphics[width=4.0in,clip]{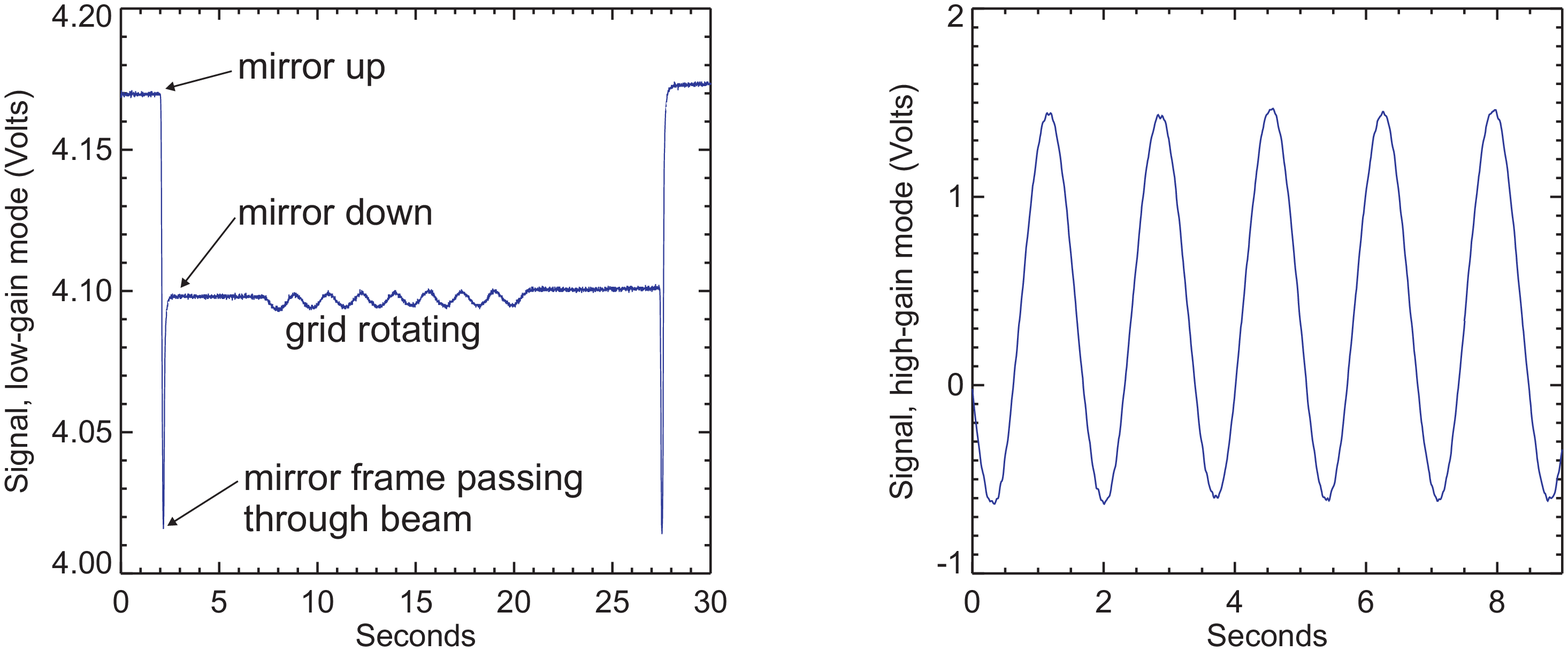}
\end{center}
\caption{(left) The calibration source is located inside the foam cone, above the secondary mirror.
When activated via the infrared link, a flip mirror reflects the signal from a rotating polarizing
grid into the telescope beam through the small hole in the secondary.
To prevent the calibration signal from saturating the detectors, the flip mirror is actually
a small aluminum disk suspended by Kevlar thread within an aluminum frame.
(middle) An early test run of the calibration source shows the procedure.
(right) A cal source run showing the same channel, taken with the receiver electronics
in the usual ``high-gain'' mode used during observation.
The flip mirror area was doubled since the test run to increase the calibration signal.}
\label{fig:calsrc_cartoon}
\vspace{0.1in}
\end{figure*}

The $100^\circ$ sidelobe was studied in Austral
summer 2007 with a series of runs during which it was deliberately scanned across
the sun while the telescope rotated through the full 360 degrees
about the boresight (deck rotation).
These were intended to precisely map the variation
of space angle with deck, but due to differing weather
conditions in summer and winter, we could not use
these results to provide tighter cuts on the CMB data.
These runs did, however, confirm that the sidelobe
is extremely narrow (FWHM $\sim 0.22^\circ$).

Since it is the result of reflection off a dielectric surface,
the sidelobe is highly polarized. Laboratory tests using a
foam cone structure identical to that on the telescope measured
the fractional polarization to be 50\% at 150\,GHz, with a
reflection coefficient of approximately 2\%. The amplitude of the electric
field of a reflected ray in the radial direction to the cone
axis of symmetry is suppressed relative to that in the
tangential direction. A PSB with its sensitivity axis aligned
with the cone tangent will therefore measure a larger
signal than a perpendicular PSB, resulting in a false polarized
signal in the differenced data. We can predict the sign of the
polarized signal if we know the PSB pair orientations, and
the angle at which a bright source appears in the frame of the
focal plane (Figure \ref{fig:pairdiff_suntod}).
The amplitude of the moon contamination signal ($\sim 3$\,mK) is
consistent with these laboratory foam cone transmission measurements,
assuming a moon brightness temperature of order 300\,K.

\section{Calibration}  \label{sec:calibration}
The QUaD calibration process comprises frequent relative calibrations,
and an absolute calibration of the final maps based on comparison with
CMB temperature measurements from other experiments.
The primary method of channel-to-channel relative gain calibration
is to nod the telescope in elevation by 1$^\circ$
(known as an ``el-nod'').
The resulting change in air mass generates an approximately linear signal of order a
Volt in the amplified time stream of each detector.
The measured signal is used to normalize the gain of each detector
to the mean gain of all detectors in a band, and includes a
correction for the elevation offset of the feeds in the focal plane at
the current telescope rotation angle.
Figure~\ref{fig:elnod} shows an example of an el-nod and the measured
distribution of relative detector gains.
During routine CMB observations, el-nods are performed during a ``cal set'' that
precedes each scan set, approximately every 30 minutes.

Each cal set also includes a run of the calibration source, which is
located above the secondary mirror, inside the foam cone (Figure
\ref{fig:calsrc_cartoon}). When in use, a flip mirror reflects the portion
of the beam passing through the secondary hole onto a rotating
polarizing grid in front of an ambient-temperature black body target,
introducing a sinusoidally modulated signal into the detector time streams.
To avoid running wires along the foam cone which could introduce
spurious polarized signals, the calibration source is battery powered
and controlled via an infrared link.
The calibration source does not uniformly illuminate the focal plane,
and hence cannot be used to correct for relative channel-to-channel
gain variation -- this is done with the el-nods.
Instead, the calibration source is used to correct for drifts in the mean
receiver gain over time, an effect
the el-nods alone cannot correct since gain drifts and changes
in atmospheric loading both effect the el-nod Volts / airmass in the same way.

The absolute calibration for QUaD is derived by cross-correlating our
CMB temperature maps with those of other experiments and is discussed in
more detail in \clempaper.
The calibration values are 0.55\,K/V at 100\,GHz and 0.47\,K/V at 150\,GHz.
These values can be estimated based on the measurements
of instrument parameters as described in Section~\ref{sec:performance}.
The Volts to Kelvin calibration factor is given by
\newcommand{\SKV}{s_{\mbox{\scriptsize K}/\mbox{\scriptsize V}}}
\newcommand{\SDC}{S_{\mbox{\scriptsize DC}}}
\be
\SKV = \left( \frac{dQ}{dT} \; \cdot \; \SDC \right)^{-1}
\ee
where the function $Q(T)$ is the power absorbed by a detector from a
beam-filling blackbody source.  This in turn is given by
\be
 Q(T) = \frac{1}{2}\trans \eta_T \int \! A \Omega \, f_n(\nu)\:B_\nu(T)\:d\nu
 \label{eq_def_q}
\ee
where $\trans$ is the atmospheric transmission (Section
\ref{sec:loading}), $\eta_T$ is the optical efficiency of the
\emph{telescope} optics, $f_n(\nu)$ is the absolute spectral bandpass
of the receiver including the optical efficiency of the receiver, $A
\Omega = \lambda^2$ is the throughput, and $B_\nu(T)$ is the Planck
function. The detector responsivity, $\SDC$, in units of V/W, is
discussed in Section \ref{sec:loadcurves}.

For the QUaD bands, the derivative $dQ/dT$ evaluated at $\Tcmb$ can be
calculated to a few percent by evaluating $dB_\nu/dT$ at the band
center and approximating the integral as
\be
 \frac{dQ}{dT} \approx
 \frac{1}{2}\trans \eta \lambda_0^2 \, \Delta\!\nu \, \left.\frac{dB_\nu}{dT}\right|_{T=\Tcmb,\:\nu=\nu_0}
 \label{eq_def_dq_dt}
\ee
where $\eta$ is the total optical efficiency of the system. The
telescope efficiency is estimated to be $\sim 0.82$ including loss from
the foam cone, the primary mirror, and most significantly shadowing
by the secondary mirror. Using average values of the
receiver properties at each frequency, this results in the calibration
estimates of
\begin{eqnarray}
\SKV(\mbox{\footnotesize 100 GHz}) &\:=\:&
     0.51 \frac{\mbox{K}}{\mbox{V}}  \times     \left(\frac{0.94}{\trans}\right) \left(\frac{0.82 \cdot 0.26}{\eta_T\,\eta_R}\right)\times\nonumber\\
  &&      \left(\frac{26.5\;\mbox{GHz}}{\Delta \nu}\right)   \left(\frac{3.6\times 10^8 \; \mbox{V}/\mbox{W} }{S_\ns{DC}}\right) \\
  &&\nonumber\\
\SKV(\mbox{\footnotesize 150 GHz}) &\:=\:&
     0.44 \frac{\mbox{K}}{\mbox{V}}  \times     \left(\frac{0.96}{\trans}\right) \left(\frac{0.82 \cdot 0.32}{\eta_T\,\eta_R}\right)\times\nonumber\\
  &&      \left(\frac{40.5 \;\mbox{GHz}}{\Delta \nu}\right)   \left(\frac{3.1\times 10^8 \;\mbox{V}/\mbox{W}}{S_\ns{DC}}\right)
\end{eqnarray}
which are in good agreement with the measured calibration values given above.

\section{Sensitivity}  \label{sec:sensitivity}
\begin{figure}[t]
   \begin{center}
\includegraphics[width=3.0in,clip]{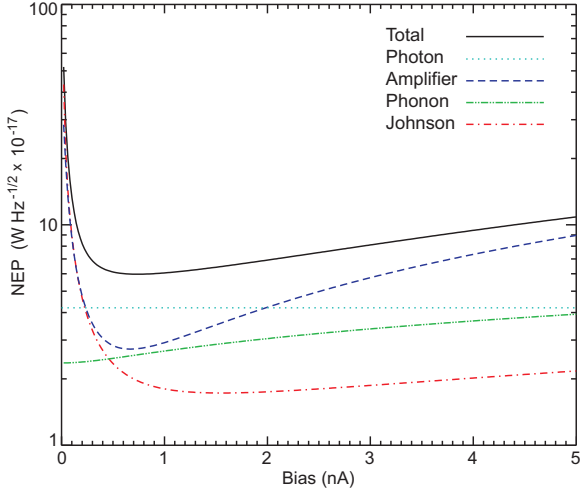}
   \end{center}
   \caption{Contributions to the NEP as a function of electrical bias current for a typical 150 GHz detector.}
   \label{fig:nep_breakdown}
\end{figure}
\begin{figure}
   \begin{center}
\includegraphics[width=3.25in,clip]{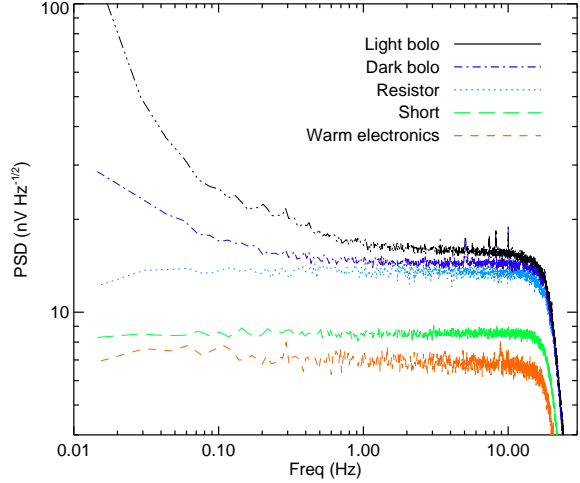}
   \end{center}
   \caption{PSD of the noise from various sources.
   ``Warm electronics'' accounts for both the readout and the bias generator and was made by connecting a warm 1\,k$\Omega$ resistor in place of the cold bolometers and amplifiers.
   ``Short'' is a measurement of the noise across a short located on the focal plane and includes contributions from the cold JFETS and the warm amplifiers (but not the bias generator).
   ``Resistor'' is a properly-biased 10~M$\Omega$ resistor on the focal plane.
   ``Dark bolo'' is the readout from a dark bolometer.
   ``Light bolo'' is the readout from a 150 GHz bolometer made during normal operation and includes contributions from atmospheric 1/f.}
   \label{fig:electronics_noise}
\end{figure}
\begin{table}
    \caption{Loading, Noise Budget, and Sensitivity}\label{tab:noise_budget}
    \centering
    \begin{tabular}{lcc}
     \hline \hline
        Frequency band (GHz)                         & 100 & 150 \\
\hline
\\
        \multicolumn{3}{l}{\em Loading (Temperatures are Raleigh-Jeans)}\\
        Total loading per detector (pW)       & 2.8  & 5.1  \\
        Total loading per detector (K)        & 27   & 26  \\
        Atmospheric loading (K)               & 11   & 9  \\
        Telescope loading (K)                 & 8    & 8  \\
        CMB loading (K)                       & 1.1  & 0.6 \\
        \\
        \multicolumn{3}{l}{\em Bolometer parameters}\\
        $R_0$    ($10^6$ Ohms)                & 94.1    & 94.8    \\
        $\Delta$ (K)                          & 42.0    & 42.1    \\
        $T_\ns{Bolo}$ (K)                     & 0.356   &  0.357   \\
        $G$ (pW / K)                          & 147     & 156    \\
        Responsivity ($10^8$ V/W)             & 3.6     & 3.1 \\
        \\
        \multicolumn{3}{l}{{\em Noise Equivalent Power} ($10^{-17}$ W/Hz$^{1/2}$)}  \\
        NEP amplifier           & 2.5 & 3.0 \\
        NEP Johnson             & 1.4 & 1.7 \\
        NEP Phonon              & 2.5 & 2.7 \\
        NEP photon shot         & 1.9 & 3.2 \\
        NEP photon Bose         & 1.7 & 2.5 \\
        Total NEP without Bose  & 4.3 & 5.4 \\
        Total NEP with Bose     & 5.6 & 5.9 \\
        NEP achieved            & 4.7 & 5.8 \\
        \\
        \multicolumn{3}{l}{{\em Sensitivity}}\\
        Calibration (K / V)                   & 0.55 & 0.47 \\
        NET per feed ($\mu$K s$^{-1/2}$)     & 440 & 390 \\
        NEQ per feed ($\mu$K s$^{-1/2}$)     & 510 & 450 \\
     \hline \hline
    \end{tabular}
     \tablecomments{Values are for a good observing day at the field elevation of $\sim 50^\circ$. See Section~\ref{sec:sensitivity}.}
\end{table}
\begin{figure}
\begin{center}
\includegraphics[width=3.25in,clip]{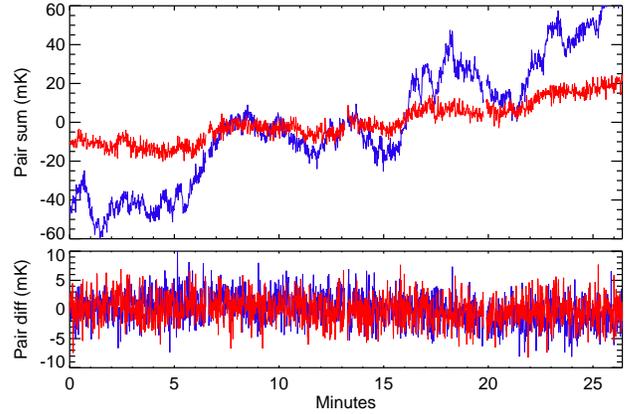}
\end{center}
\caption{Time ordered data for a 150\,GHz detector pair (blue) and a 100\,GHz pair (red) while scanning the telescope during routine CMB observation.  Atmospheric $1/f$ noise, readily apparent in the pair sum (top) is nearly all common mode and cancels upon differencing (bottom). \label{fig:quad_tod}}
\end{figure}
\begin{figure}
\begin{center}
\parbox{3.6in}{\hspace{-0.25in}
\includegraphics[width=3.8in,clip]{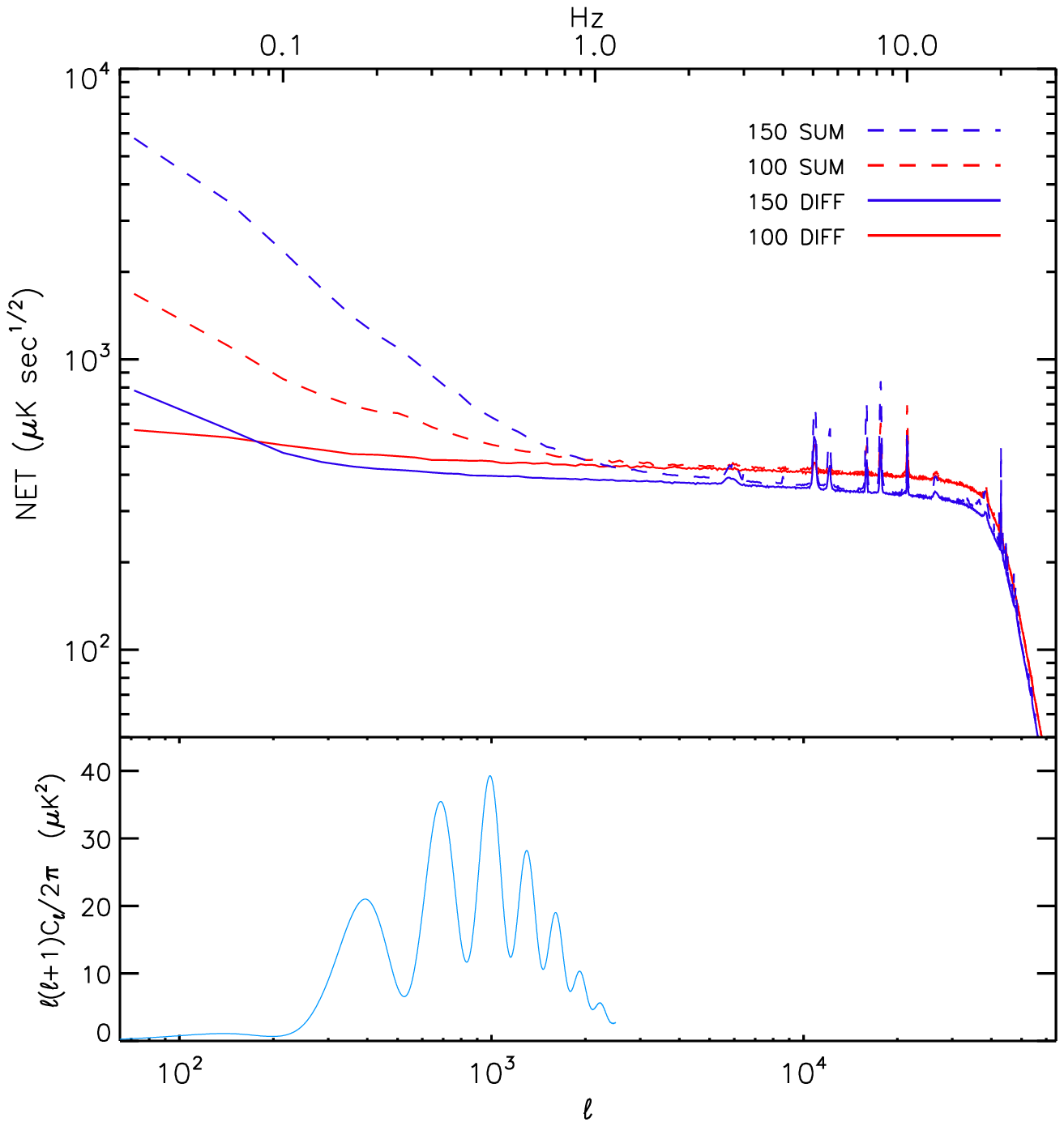}
}
\end{center}
\caption{Average per channel NET for PSB sum (dashed) and difference (solid) signals from data taken on Jun 23, 2006. The telescope
was scanning at 0.17 deg/sec on the sky. The E-mode CMB power spectrum is shown below
assuming this scanning speed.  The spikes seen in the data around approximately 10\,Hz are microphonic resonances. For analysis, the data is low-pass filtered in software. Note that above a frequency of approximately 2.5\,Hz, the sensitivity is reduced due to the detector time constants and the antenna beam size (not indicated on this plot). \label{fig:psd_net}}
\end{figure}
The sensitivity of bolometric detectors is limited by Johnson noise
from the thermistor and load resistors, phonon noise due to the
quantization of the heat conducted between the absorber and the bath,
and amplifier noise generated in the readout electronics
\citep{mather1,mather2}. Additionally there are two sources of noise
associated with the quantum nature of the incident optical radiation;
shot noise from the discrete photon arrival times and ``clumping''
noise due to the Boson properties of photons. See, for example,
Appendix B of \cite{runyan2003} for a thorough discussion of this
second term.

For millimeter-wave systems, the detector sensitivity is usually quoted
as a noise equivalent power (NEP) in W\,Hz$^{-1/2}$, computed by dividing the power spectrum in Volts by the detector
responsivity (Volts / Watt). Figure~\ref{fig:nep_breakdown} shows the expected contributions
to the NEP for a typical QUaD detector based on the measured properties
of the bolometers and readout electronics. Figure~\ref{fig:electronics_noise} shows the measured noise contributions from
the readout electronics.

The sensitivity to CMB temperature fluctuations is quantified as a noise equivalent
temperature (NET), which is computed by multiplying the detector
voltage noise by the absolute calibration factor in units of Kelvin / Volt (Section~\ref{sec:calibration}).
The NET of a detector is related to the measured
NEP as
 \be
 \NET{} = \frac{\NEP{}} {\sqrt{2}\,\left.(dQ/dT)\right|_{\Tcmb}}
 \ee
where the factor of $\sqrt{2}$ in the denominator converts from units of Hz$^{-1/2}$ to
sec$^{1/2}$ and $dQ/dT$ is defined in (\ref{eq_def_dq_dt}).
This form is frequently used for estimating the sensitivity of a system during the design
phase.
Similarly, the NEQ gives the sensitivity to CMB polarization anisotropies.
Since polarization is measured by differencing orthogonal PSB pairs, common mode
atmospheric noise is greatly reduced (Figure~\ref{fig:quad_tod}).
However, the NEQ is degraded relative to the NET by a factor of $(1+\epsilon)/(1-\epsilon)$
(see Equation~\ref{eq_v_psb}) which for QUaD is $\sim 1.16$ due to cross-polar leakage.

Figure \ref{fig:psd_net} shows the sensitivity achieved during a typical CMB observing day.
The PSD is estimated by averaging the Fourier transform of each of the approximately one thousand azimuth scans performed during that day.
Each scan is approximately $7^\circ$ on the sky and takes 40 seconds.
Two PSDs are computed for each PSB pair, one by summing the data, and the other by subtracting.
The pair sum and difference PSDs are averaged over all the operational channels in each frequency band.
Only minimal filtering is applied to the raw data --- a best fit line is removed from each half scan.

The average NET for a feed horn (PSB pair), not including the effects of atmospheric $1/f$ noise, is computed
from the pair \emph{difference} data as 440 and 390\,$\mu$K\,$\sqrt{\mbox{sec}}$ at 100 and 150\,GHz, respectively over the science band (0.1 to 1.0\,Hz).
The NET for the two bands, including the effects of atmospheric $1/f$, is computed from the pair \emph{sum} PSD to be 490 and 580\,$\mu$K\,$\sqrt{\mbox{sec}}$.
Since the atmosphere adds a large amount of $1/f$-noise, the NET achieved over the course of a season depends on the angular scale of interest,
on the filtering used during analysis, and on the weighting used to combine data taken during differing weather conditions.
The NEQ for QUaD can be estimated by multiplying the NET computed from the PSB pair difference data by the polarization efficiency factor and is found to be 510 and 450\,$\mu$K\,$\sqrt{\mbox{sec}}$.
Table~\ref{tab:noise_budget} gives the noise budget and sensitivity for QUaD.

\section{Conclusion}  \label{sec:conclusion}
We have presented the design and performance of the QUaD experiment.
Stability of key instrumental parameters such as the pointing, beams,
and polarization efficiencies have been shown to be
at a level that does not interfere with the data collection.
Sources of data contamination, including atmospheric noise
and sidelobe pickup, are sufficiently well understood to be
dealt with effectively --- a combination of filtering, field differencing, and data cuts suppresses
them to below the level of the instrument sensitivity.
QUaD began data collection in early 2005 and completed three seasons
of observation before being decommissioned in late 2007.
Analysis of these surveys is presented in \citet{quad_season1} and \citet{Pryke2008}.

\vspace{0.15in}
We acknowledge the staff of the Amundsen-Scott South Pole Station and all involved in the United States Antarctic Program for their superb support during the construction and operation of this experiment.
Special thanks go to our brave winter-over scientist Robert Schwarz who has spent three consecutive winter seasons with the
QUaD.
We would also like to acknowledge the tremendous efforts of the Stanford University Physics Department machine shop in the construction of the focal plane assembly.
JRH would like to thank David Chuss for useful comments on this draft
and Simon Radford for providing the 350\,$\mu$m tipper data.
QUaD is funded by the National Science Foundation in the USA, through grants AST-0096778,ANT-0338138,ANT-0338335 and ANT-0338238, by the UK Science and technology Facilities Council (STFC) and its predecessor the Particle Physics and Astronomy Research Council (PPARC), and by the Science Foundation Ireland.

JRH acknowledges the support of an NSF Graduate Research
Fellowship, a Stanford Graduate Fellowship, and a NASA Postdoctoral Fellowship.
MLB and AO acknowledge the award of PPARC fellowships.
PGC is funded by the {\it Funda\c{c}\~ao para a Ci\^encia e a Tecnologia}.
SEC acknowledges support from a Stanford Terman Fellowship.
JMK acknowledges support from a John B. and Nelly L. Kilroy Foundation Fellowship.
CP and JEC acknowledge partial support from the Kavli Institute
for Cosmological Physics through the grant NSF PHY-0114422.
EYW acknowledges receipt of an NDSEG fellowship.
MZ acknowledges the support of a NASA Postdoctoral Fellowship.
This research was supported in part by appointments to the NASA
Postdoctoral Program at the Goddard Space Flight Center (JRH) and the Jet
Propulsion Laboratory (MZ), administered by Oak Ridge
Associated Universities through a contract with NASA. 

\newpage
\bibliographystyle{apj}
\bibliography{instrument}
\end{document}